\documentclass[12pt,authoryear,notitlepage]{article}\usepackage[top=25mm,bottom=25mm,left=25mm,right=25mm]{geometry}
\usepackage[hidelinks]{hyperref}

\usepackage{array}
\usepackage{booktabs,dcolumn}
\usepackage[table]{xcolor}
\usepackage{mathptmx}       
\usepackage{helvet}         
\usepackage{courier}        
\usepackage{type1cm}        
\usepackage{makeidx}         
\usepackage{graphicx}        
\usepackage{multicol}        
\usepackage[bottom]{footmisc}
\usepackage[flushleft]{threeparttable}
\usepackage{adjustbox}
\usepackage{natbib}
\usepackage{placeins}
\usepackage{url}                         
\PassOptionsToPackage{obeyspaces}{url}
\makeindex             

\usepackage{amsmath,amsfonts,amssymb,amsthm}

\theoremstyle{plain}

\newtheorem{theorem}{Theorem}

\usepackage{lipsum}
\usepackage{enumitem}

\setlist[itemize]{leftmargin=*}
\newcommand{\virg}[1]{``#1''}
\usepackage{caption}
\usepackage{setspace}
\usepackage{lscape}
\usepackage{arydshln}
\usepackage{float}
\usepackage{tabularx}
\usepackage{multirow}

\usepackage{bm}
\DeclareMathOperator*{\argmin}{arg\,min} 

\usepackage{enumitem}
\usepackage{rotating}

\setlist[itemize]{leftmargin=*}

\newcolumntype{.}{D{.}{.}{-1}}
\makeatletter
\newcolumntype{B}[3]{>{\boldmath\DC@{#1}{#2}{#3}}c<{\DC@end}}
\makeatother

\newcommand\mc[1]{\multicolumn{1}{c}{#1}} 
\newcolumntype{d}{D{.}{.}{5.5}}           
\newcolumntype{p}{D{.}{}{5.5}}  

\usepackage{pdflscape}

\usepackage{ragged2e}

\usepackage{tipa}

\usepackage{subcaption}
\usepackage{bbm}
\usepackage[cal=cm]{mathalfa}

\usepackage{authblk}

\title{Mixed--frequency quantile regressions to forecast Value--at--Risk and Expected Shortfall\thanks{A previous draft was circulated under the title \emph{Mixed--frequency quantile regression with realized volatility to forecast Value-at-Risk}}}

\author[a]{Vincenzo Candila\thanks{\href{mailto:vcandila@unisa.it}{vcandila@unisa.it} (Corresponding author)
}
}

\author[b]{Giampiero M. Gallo\thanks{\href{giampiero.gallo@nyu.edu}{giampiero.gallo@nyu.edu}
}}

\author[c]{Lea Petrella\thanks{\href{mailto:lea.petrella@uniroma1.it}{lea.petrella@uniroma1.it}
}
}

\affil[a]{Department of Economics and Statistics, University of Salerno, Italy}
\affil[b]{Italian Court of Audits (Corte dei conti) and NYU in Florence}
\affil[c]{Department of Methods and Models for Economics, Territory and Finance, Sapienza University of Rome, Italy}

\providecommand{\keywords}[1]
{
  \small	
  \textbf{Keywords:} #1
}

\providecommand{\jel}[1]
{
  \small	
  \textbf{JEL:} #1
}

\begin{document}
\maketitle

\begin{abstract}
{\footnotesize 
Although quantile regression to calculate risk measures has been widely established in the financial literature, when considering data observed at mixed--frequency, an extension is needed. In this paper,  a model is suggested built on a mixed--frequency quantile regression to directly estimate the Value--at--Risk (VaR) and the Expected Shortfall (ES) measures. In particular, the low--frequency component incorporates information coming from variables observed at, typically, monthly or lower frequencies, while the high--frequency component can include a variety of daily variables, like market indices or realized volatility measures. The conditions for the weak stationarity of the daily return process are derived and the finite sample properties are investigated in an extensive Monte Carlo exercise. The validity of the proposed model is then explored through a real data application using two energy commodities, namely, Crude Oil and Gasoline futures. Results show that our model outperforms other competing specifications, on the basis of some popular VaR and ES backtesting test procedures.
}			
\end{abstract}
		
\\

\jel{C22, C52, C53, C58.}

\section{Introduction}

Risk management has spurred a vast literature in financial econometrics to meet the challenges imposed by the Basel--II and Basel--III agreements and develop model--based approaches to calculate regulatory capital requirements \citep{Kinateder:2016} in a forecasting perspective. For \textit{tail market risk}, special attention was devoted to the Value--at--Risk (VaR) measure at a given confidence level $\tau$, VaR($\tau$), defined as the worst portfolio value movement (return) to be expected at $1-\tau$ probability over a specific horizon \citep{Jorion:1997}. The VaR measure is complemented by another tail risk measure called Expected Shortfall (ES), defined as the conditional expectation of returns in excess of the VaR \citep[see][among others]{Acerbi:Tasche:2002a,Rockafellar:Uryasev:2002}. Unlike VaR, ES is a coherent risk measure \citep{Artzner_et_al:1999,Acerbi:Tasche:2002b} and provides deeper information on the shape and the heaviness of the tail in the loss distribution. Together, such measures represent the most popular benchmark in the risk management practice \citep{Christoffersen:Goncalves:2005,Sarykalin_et_al:2008}.

Being the $\tau$--quantile of a portfolio return distribution, the VaR($\tau$) can be predicted as the product of the portfolio volatility forecast times the quantile of the hypothesized distribution. For the first component, volatility clustering, modeled by conditionally autoregressive models \citep[such as the ARCH/GARCH -- ][]{Engle:1982, Bollerslev:1986}, produces good forecasts capable of reproducing well known stylized facts of financial time series, including skewed behavior and fat tails \citep[][among others]{Cont:2001,Engle:Patton:2001}. Further improvements were made possible by the direct predictability of realized measures of financial volatility \citep{Andersen:Bollerslev:Christoffersen:Diebold:2006}. While a choice of a specific parametric distribution for the innovation term may be uninfluential for model parameter estimation  \citep{Bollerslev:Wooldridge:1992}, unless a few extreme events  \citep[e.g. the Flash Crash of May 2005 or the presence of outliers,][]{Carnero:Pena:Ruiz:2012} occur, a wrong choice of distribution for the innovation term delivers inaccurate quantiles and hence an inadequate VaR($\tau$) forecasting: see for example \cite{Manganelli:Engle:2001} and \cite{Ghourabi:Francq:Telmoudi:2016}.


As an alternative, the VaR($\tau$) can be directly derived through quantile regression methods \citep{Koenker:Bassett:1978, Engle:Manganelli:2004} where no distributional hypothesis is required. A first suggestion in this direction comes from \cite{Koenker:Zhao:1996} who use quantile regression for a particular class of ARCH models, i.e., the Linear ARCH models \citep{Taylor:1986}, chosen for its ease of tractability in deriving theoretical properties. Subsequent refinements are, for instance,  \cite{Xiao:Koenker:2009}, \cite{Lee:Noh:2013}, \cite{Zheng:Zhu:Li:Xiao:2018} for GARCH models,  \cite{Noh:Lee:2016} who consider asymmetry, \cite{Chen:Gerlach:Hwang:McAleer:2012} who consider nonlinear regression quantile approach with intra-day price, \cite{Bayer:2018} who combines VaR forecasts via penalized quantile regressions, \cite{Taylor:2019} who considers the Asymmetric Laplace distribution to jointly estimate VaR and ES and the multivariate generalization of \cite{Merlo:Petrella:Raponi:2021}. 

A relatively recent stream of literature investigates the value of information provided by data available at both high-- and low--frequency incorporated into the same model in assessing the dynamics of financial market activity: this is the case of the GARCH--MIDAS model proposed by \cite{Engle:Ghysels:Sohn:2013} \citep[building on the MI(xed)--DA(ta) Sampling approach by][]{Ghysels:Sinko:Valkanov:2007}, the regime switching GARCH--MIDAS of \cite{Pan:Wang:Wu:Yin:2017}, the recent paper by \cite{Xu:Wang:Liu:2021} who consider a MIDAS component in the Conditional Autoregressive Value--at--Risk (CAViaR) of \cite{Engle:Manganelli:2004},  the work of  \cite{Pan:Wang:Liu:2021} where the parameters of the GARCH-MIDAS models for jointly calculating VaR and ES are obtained through the loss function of \cite{Fissler:Ziegel:2016}, and the contribution of \cite{Xu_et_al:2022} who calculate the weekly tail risks of three market indices using information from daily variables. 

The main contribution of this paper is a novel Mixed--Frequency Quantile Regression model \citep[MF--QR,  extending][]{Koenker:Zhao:1996}: we show how the constant term in the quantile regression can be written as a function of data sampled at lower frequencies (and hence become a low--frequency component), while the high--frequency component is regulated by the daily data. As a result, with the aim of capturing dependence on the business cycle, we benefit from the information contained in low--frequency variables \citep[cf.][among others]{Mo:Gupta:Li:Singh:2018,Conrad:Loch:2015}, and we achieve a rather flexible representation of volatility dynamics. Since both components enter additively, our model can be seen as a quantile model version of the Component GARCH by \cite{Engle:Lee:1999}.



In the proposed model, we also include a predetermined variable observed daily, typically a realized measure: this adds the \virg{--X} component in the resulting MF--QR--X model. This variable can capture extra information useful in modeling and forecasting future volatility and may improve the accuracy of tail risk forecasts. Such a use in the quantile regression framework is not new in itself: the paper by \cite{Gerlach:Wang:2020} jointly forecasts VaR and ES and \cite{Zhu:Li:Xiao:2020} predict VaR by adopting a GARCH--X model for the volatility term. Also the work of \cite{Zikes:Barunik:2016} uses the realized measures in the context of quantile regressions to investigate the features of conditional quantiles of realized volatility and asset returns.

The proposed MF--QR--X specification and its nested alternatives \citep[including, the QR version of][]{Koenker:Zhao:1996} belong to the class of semi-parametric models, without resorting to  restrictive assumptions about the error term distribution and are able to calculate the VaR directly. Such a model can also jointly forecast the VaR and ES via the Asymmetric Laplace distribution as proposed by \cite{Taylor:2019}.

From a theoretical point of view, we provide the conditions for the weak stationarity of the daily return process suggested. The finite sample properties are investigated through an extensive Monte Carlo exercise. The empirical application is carried out on the VaR and ES predictive capability for two energy commodities, the West Texas Intermediate (WTI) Crude Oil\footnote{The VaR and ES of this commodity have been recently investigated by \cite{Kuang:2022}} and the Reformulated Blendstock for Oxygenate Blending (RBOB) Gasoline futures, both observed daily.  
The period under investigation  starts on January 2010 and ends on July 2022, covering both the Covid-19 pandemic and some consequences of the Russian aggression of Ukraine. The competing models consist of many common parametric, semi--parametric and non--parametric choices. Some parametric models like the GARCH--MIDAS use the same low--frequency variable employed in the proposed MF-QR--X specification. Given our empirical interest in evaluating risks related to energy commodities, a relevant choice for such a variable is the geopolitical risk (GPR) index proposed by \cite{Caldara:Iacoviello:2022}, observed monthly.\footnote{The monthly GPR index we use is built through an automated text-search on the articles of ten newspapers in relationship to eight risk categories. Such an  index has been extensively used in many recent contributions concerning oil volatility \citep[see, for instance,][among others]{Liu_et_al:2019,Mei_et_al:2020,Qin_et_al:2020}.} The resulting VaR and ES predictions are evaluated in-- and out--of--sample, according to the customary backtesting procedures: our out--of--sample period starts on January 2017 and ends on July 2022, and the VaR and ES forecasts are obtained using a rolling window that updates the parameter estimates every five, ten and twenty days. The results show that our MF--QR--X outperforms all the other competing models considered, proving the merits of resorting to a mixed--frequency source of information.
The useful contribution of a low--frequency variable in a risk management perspective thus lies in its capability of capturing secular movements in the conditional distributions related to risk factors slowly shifting through time. 

The rest of the paper is organized as follows. In Section \ref{sec:VaR} we introduce the notation and the basis for a dynamic model for the VaR and ES and we provide details of the conditional quantile regression approach. Section \ref{sec:ourmodel} presents our MF--QR--X model.  Section \ref{sec:monte_carlo} is devoted to the Monte Carlo experiment. Section \ref{sec:back_model_selec} details the backtesting procedures. Section \ref{sec:emp_analysis} illustrates the empirical application. Conclusions follow.

\section{Approaches to VaR and ES estimation \label{sec:VaR}}

For the purposes of this paper we will adopt a double time index, $i,t$, where $t=1,\ldots,T$ scans a low frequency time scale (i.e., monthly) and $i=1,\ldots,N_t$ identifies the day of the month, with a varying number of days $N_t$ in the month $t$, and an overall number $N$ of daily observations $N=\sum_{t=1}^T N_t$. Let the daily returns $r_{i,t}$ be, as customarily defined, the log--first differences of prices of an asset or a market index, and let the information available at time $i,t$ be $\mathcal{F}_{i,t}$.  
In what follows, we are interested in the conditional distribution of returns, with the assumption: 
\begin{equation}
	r_{i,t} = \sigma_{i,t} z_{i,t} \quad \text{with} \quad t=1,\ldots,T, \ i=1,\ldots,N_t,
	\label{eq:cond_dist}
\end{equation}
where $z_{i,t} \overset{iid}\sim (0,1)$ having a cumulative distribution function denoted by $F(\cdot)$.
The zero conditional mean assumption in Eq. (\ref{eq:cond_dist}) is not restrictive, in fact, when explicitly modeled, such a conditional mean is very close to zero, consistently with the market efficiency hypothesis. 

Based on this setup, the conditional (one-step-ahead) VaR for day $i,t$  at $\tau$ level ($VaR_{i,t}(\tau)$) for $r_{i,t}$ is defined as
$$
Pr(r_{i,t} < VaR_{i,t}(\tau)|\mathcal{F}_{i-1,t})= \tau,
$$
i.e., the $\tau$-th conditional quantile of the series $r_{i,t}$, given $\mathcal{F}_{i-1,t}$; consequently, we can write 
\begin{equation}
VaR_{i,t}(\tau)\equiv Q_{r_{i,t}}\left(\tau|\mathcal{F}_{i-1,t}\right)= \sigma_{i,t} F^{-1}(\tau),
\label{eq:VaR}
\end{equation} 
where ${F^{-1}(\tau)}= \inf \left\lbrace z_{i,t}: F(z_{i,t}) \geq \tau \right\rbrace$. For a given $\tau$, the traditional volatility--quantile approach to estimate the $VaR_{i,t}(\tau)$ is thus based on modeling $\sigma_{i,t}$ from a dynamic model of either the conditional variance of returns \citep[following][]{Engle:1982, Bollerslev:1986} or as a conditional expectation of a realized measure \citep{Andersen:Bollerslev:Christoffersen:Diebold:2006b} and retrieving the constant $F^{-1}(\tau)$ either parametrically or nonparametrically.
In either case, from an empirical point of view, it turns out that distribution tests mostly reject specific parametric choices, and that using the empirical distributions is prone to bias/variance problems and lack of stability through time.

Alternatively, we can estimate directly $Q_{r_{i,t}}\left(\tau|\mathcal{F}_{i-1,t}\right)$ using a quantile regression approach \citep{Koenker:Bassett:1978, Engle:Manganelli:2004} which has become a widely used technique in many theoretical problems and empirical applications. While classical regression aims at estimating the mean of a variable of interest conditioned to regressors, quantile regression provides a way to model the conditional quantiles of a response variable with respect to a set of covariates in order to have a more robust and complete picture of the entire conditional distribution. This approach is quite suitable to be used in all the situations where specific features, like skewness, fat--tails, outliers, truncation, censoring and heteroskedasticity are present. The basic idea behind the quantile regression approach, as shown by \cite{Koenker:Bassett:1978}, is that the $\tau$-th quantile of a variable of interest (in our case $r_{i,t}$), conditional on the information set $\mathcal{F}_{i-1,t}$, can be directly expressed as a linear combination of a $q+1$ vector of variables ${x}_{i-1,t}$ (including a constant term),  with parameters ${\Theta}_\tau$, that is: 
\begin{equation}
	Q_{r_{i,t}}\left(\tau|\mathcal{F}_{i-1,t}\right) = {x}_{i-1,t}' {\Theta}_\tau.
	\label{eq:quant}
\end{equation} 
An estimator for the $(q+1)$ vector of coefficients  ${{\Theta}}_\tau$ is obtained minimizing a suitable loss function (also known as check function):
\begin{equation}
\hat{{\Theta}}_\tau =	\argmin_{{\Theta}}  \sum  \rho_{\tau} \left( r_{i,t} - {x}_{i-1,t}' {\Theta_\tau} \right),
	\label{eq:check_fun_rho}
\end{equation}
with $\rho_\tau(u)=u\left( \tau - \mathbbm{1}\left(u<0\right)\right)$, where $\mathbbm{1}\left(\cdot\right)$ denotes an indicator function. In our context, the advantage of such an approach  is to avoid the need to specify the distribution of $z_{i,t}$ in Eq. (\ref{eq:cond_dist}), either parametrically or nonparametrically.

Following the approach by \cite{Koenker:Zhao:1996}, we assume a dependence of $\sigma_{i,t}$ on past absolute values of returns:
\begin{equation}
\sigma_{i,t} =\beta_0 + \beta_1 |r_{i-1,t}| + \ldots + \beta_q |r_{i-q,t}|, \quad \text{ with } \quad t=1,\ldots,T, \ i=1,\ldots,N_t,
\label{eq:lin_arch}
\end{equation} 
with  $0<\beta_0<\infty$, $\beta_1,\ldots,\beta_q\geq 0$. Thus, substituting  the generic term ${x}_{i-1,t}$ in Eq. (\ref{eq:quant}) with the specific vector in Eq. \eqref{eq:lin_arch}, we have
\begin{equation}
\sigma_{i,t}=\left(1, |r_{i-1,t}|,\ldots,|r_{i-q,t}|\right)'\left(\beta_0,\beta_1,\ldots,\beta_q\right) = x_{i-1,t}'\Theta.
\label{eq:x_larch}
\end{equation} 
Such an approach turns out to be convenient, since it allows for a direct comparability of the two setups to estimate the VaR($\tau$) in Eq. (\ref{eq:VaR}):
\begin{equation}
VaR_{i,t}(\tau) = \left\lbrace \begin{array}{ll}
	x_{i-1,t}'\Theta \ F^{-1}(\tau)&\textrm{volatility--quantile} \\
	x_{i-1,t}'\Theta_\tau&\textrm{conditional quantile regression},
\end{array}
\right.
\label{eq:var_quant_vol}
\end{equation}
 
\noindent which establishes the equivalence $\Theta \, F^{-1}(\tau)=\Theta_\tau$ which will prove useful later in our Monte Carlo simulations. \textcolor{black}{Moreover, as also pointed out by \cite{Koenker:Zhao:1996}, what we estimate in the conditional quantile regression framework are the parameters in $\Theta_{\tau}$, which are different from the parameters included in $\Theta$ of the volatility--quantile context. While the  parameters in $\Theta$ are constrained to be non--negative, the parameters in $\Theta_{\tau}$ may be negative depending on the value of $\tau$.} The volatility--quantile and conditional quantile regression options in Eq. \eqref{eq:var_quant_vol} give rise to the so-called parametric and semi-parametric models for the VaR, respectively. Alternatively, the most prominent example of a non-parametric approach to derive the VaR is the Historical Simulation  \citep[HS --][]{Hendricks:1996}. The HS model calculates this risk measure as the empirical quantile over a window of returns with length $w$, that is:
\begin{equation}
VaR_{i,t}(\tau)={Q}_{\bm r_{i,t}^{w}}(\tau),
\label{eq:var_hs}
\end{equation}
where $\bm r_{i,t}^{w} = (r_{i-w,t},r_{i-w+1,t},\dots,r_{i-1,t})$. 

The linear representation in \eqref{eq:lin_arch} can be further justified by noting that the term $\sigma_{i,t}$ defining the volatility of returns can also be seen as the conditional expectation of absolute returns in the Multiplicative Error Model representation used by \cite{Engle:Gallo:2006}:
\begin{equation}
|r_{i,t}|=\sigma_{i,t} \eta_{i,t}.
\label{eq:MEM}
\end{equation}
The term $\eta_{i,t}$ is an i.i.d. innovation with non--negative support and unit expectation, and the Eq. (\ref{eq:MEM}) can be used to derive an estimate of the VaR. The representation in \eqref{eq:lin_arch} can  also be seen as a simple and convenient nonlinear autoregressive model for  $|r_{i,t}|$ with multiplicative errors, which we hold as the maintained base specification to explore the merits of our proposal. Moreover, this lays the grounds for extending the approach, using other specifications for $\sigma_{i,t}$ in Eq. \eqref{eq:lin_arch} as functions of past volatility--related observable variables. For example, as an alternative, we can be considered:
$$
\sigma_{i,t} =\omega + \alpha_1 rv_{i-1,t} + \ldots + \alpha_q rv_{i-q,t}, \quad \text{ with } \quad t=1,\ldots,T, \ i=1,\ldots,N_t,
$$ 
with $rv_{i,t}$ the daily realized volatility. 

A similar framework can be adopted to calculate the ES, following, again, the same parametric, non-parametric and semi-parametric approaches as before. The parametric models with Gaussian error distribution calculate the ES through:
\begin{equation}\label{es_g_dist_1}
\text{ES}_{i,t}(\tau)=  -h_{i,t}^{1/2} \frac{\phi(\Phi^{-1}(\tau))}{\tau} \,,
\end{equation}
\noindent where $h_{i,t}$ is the conditional variance, $\phi(\cdot)$ and $\Phi^{-1}(\tau)$ are the probability density function (PDF) and quantile function of the standard Gaussian distribution, respectively.
The parametric models with Student's t error distribution calculate the ES via:
\begin{equation}\label{es_t_dist_1}
\text{ES}_{i,t}(\tau)=  -h_{i,t}^{1/2}  \left( \frac{g_{\nu}(G_{\nu}^{-1}(\tau))}{\tau} \right) \left( \frac{\nu+ (G_{\nu}^{-1}(\tau))^{2}}{\nu-1} \right) \sqrt{\frac{\nu-2}{\nu}} \,,
\end{equation}
\noindent where $g_{\nu}$ and $G_{\nu}^{-1}(\tau)$ are the PDF and quantile function of the Student's t with $\nu$ degrees of freedom, respectively.

The HS calculates the ES as follows:

\begin{equation}\label{eq:essample}
{ES}_{i,t}(\tau)= \frac{\sum_{i=1}^w r_{t-w-1+i} \mathbbm{1}_{(r_{t-w-1+i} \leq {VaR}_{i,t}(\tau))}}{\sum_{i=1}^w \mathbbm{1}_{(r_{t-w-1+i} \leq {VaR}_{i,t}(\tau))}},
\end{equation}
where ${VaR}_{i,t}(\tau)$ is the VaR obtained through Eq. \eqref{eq:var_hs}.

Following \cite{Taylor:2019}, the quantile regression framework allows to jointly estimate the VaR and ES  by maximizing the following Asymmetric Laplace density (ALD), that is:

\begin{equation}
f(r_{i,t} \mid VaR_{i,t}(\tau), \tau) =  
\frac{\tau-1}{ES_{i,t}(\tau)} \exp \left( \frac{\left( r_{i,t} - {VaR}_{i,t}(\tau)\right) \left( \tau - \mathbbm{1}_{(r_{i,t} \leq {VaR}_{i,t}(\tau))}\right)}{\tau ES_{i,t}(\tau)} \right), \label{eq:aldES}
\end{equation}

\noindent where the ES in \eqref{eq:aldES} is calculated as:
\begin{equation}\label{eq:mul}
ES_{i,t}(\tau) = \left(1 + \exp\left(\gamma_{\text{ES}}\right)\right) VaR_{i,t}(\tau).
\end{equation}

We now move to the introduction of our MIDAS extension to the model in \eqref{eq:lin_arch} in a quantile regression framework, taking advantage of the well--known predictive power of low--frequency variables for the volatility observed at a daily frequency \citep[e.g.][]{Conrad:Kleen:2020}. 
We also add an \virg{--X} term to the proposed specification. This additional high--frequency variable could be a lagged realized measure of volatility \citep[see also][within a CAViAR context]{Gerlach:Wang:2020}, in order to add the informational content of a more accurate measure to the volatility dynamics, or a volatility index, like the VIX, or even accommodate asymmetric effects associated to negative returns.

\section{The MF--QR--X model \label{sec:ourmodel}}
 
\subsection{Model specification and properties}
In order to take advantage of the information coming from variable(s) observed at different frequency we introduce a low--frequency component in model (\ref{eq:lin_arch}). This low--frequency term represents a one--sided filter of $K$ lagged realizations of a given variable $MV_t$ (any low--frequency variable), through a weighting function $\delta(\omega)$, where $\omega=(\omega_1, \omega_2)$.  Our resulting Mixed--Frequency Quantile Regression (MF--QR) model becomes:
\begin{eqnarray}
r_{i,t} &=& \left[\left(\beta_0 + \theta \Big\lvert\sum_{k=1}^K \delta_k(\omega)MV_{t-k}\Big\rvert\right) + \left(\beta_1 |r_{i-1,t}| + \ldots + \beta_q |r_{i-q,t}| \right) \right] z_{i,t} \\
&\equiv& \left[(\beta_0 + \theta |WS_{t-1}|) + (\beta_1 |r_{i-1,t}| + \ldots + \beta_q |r_{i-q,t}| ) \right] z_{i,t},
\label{eq:arch_midas}
\end{eqnarray}
where the parameter $\theta$ represents the impact of the weighted summation of the $K$ past realizations of $MV_t$, observed at each period $t$, that is, $WS_{t-1}= \sum_{k=1}^K \delta_k(\omega)MV_{t-k}$. The importance of each lagged realization of $MV_t$ depends on $\delta(\omega)$, which can be assumed as a Beta or Exponential Almon lag function  \citep[see, for instance,][]{Ghysels:Qian:2019}. Here we use the former function, that is:
\begin{equation}
\delta_k(\omega)=\frac{(k/K)^{\omega_1-1} (1-k/K)^{\omega_2-1}}{\sum_{j=1}^K (j/K)^{\omega_1-1}(1-j/K)^{\omega_2-1}}.
\label{eq:beta}
\end{equation}
Eq. (\ref{eq:beta}) is a rather flexible function able to accommodate various weighting schemes. Here we follow the literature and give a larger weight to the most recent observations, that is, we set $\omega_1=1$ and $\omega_2 \geq 1$.  The resulting weights $\delta_k(\omega)$ are at least zero and at most one, and their sum equals one, so that $ \sum_{k=1}^K \delta_k(\omega)MV_{t-k}$ is an affine combination of $\left(MV_{t-1},\cdots,MV_{t-K}\right)$.

In order to refine the VaR dynamics in our model, we include a predetermined variable $X_{i,t}$, so that we can explore the empirical merits of such an extended specification, already present in the GARCH and MEM literature \citep{Han+Kristensen:2015, Engle:Gallo:2006}. Such a variable may be the realized volatility of the asset or a market volatility index \citep[see the use of the VIX in ][among others]{Amendola:Candila:Gallo:2021}. The resulting eXtended Mixed--Frequency Quantile Regression model, labelled MF--QR--X, becomes:
\begin{equation}
r_{i,t} = \left[(\beta_0 + \theta |WS_{t-1}|) + (\beta_1 |r_{i-1,t}| + \ldots + \beta_q |r_{i-q,t}| + \beta_X |X_{i-1,t}| ) \right] z_{i,t}.
\label{eq:larch_midas_x_1}
\end{equation}

In either Eq. \eqref{eq:arch_midas} or  \eqref{eq:larch_midas_x_1}, the first component (including the constant) depends only on the low--frequency term (changing at every $t$, according to the term $WS_{t-1}$), while the second comprises variables changing  daily (i.e., every ${i,t}$) and include lagged returns and the high--frequency term. In such a representation, the two components enter additively, in the spirit of the component model of \cite{Engle:Lee:1999}:
\begin{equation}
	r_{i,t} =  \left[ \sigma_t^{LF}+\sigma_{i,t}^{HF}\right] z_{i,t},
	\label{eq:larch_midas_x_two_comp}
\end{equation}
which, for the MF--QR--X model, becomes
\begin{eqnarray}
r_{i,t} &=& \left[\underbrace{(\beta_0 + \theta |WS_{t-1}|)}_{\sigma_t^{LF}} + \underbrace{(\beta_1 |r_{i-1,t}| + \ldots + \beta_q |r_{i-q,t}| + \beta_X |X_{i-1,t}| )}_{\sigma_{i,t}^{HF}} \right] z_{i,t}.
\label{eq:larch_midas_x}
\end{eqnarray}

\noindent In the following theorem we show that, under mild conditions, the process in (\ref{eq:larch_midas_x}) is weakly stationary:
\begin{theorem}
	Let $MV_{t}$ and $X_{i,t}$ be weakly stationary processes.  Assume that $\beta_0 >0$, $\beta_1,\cdots,\beta_q,\beta_x \geq 0$ and $\theta \geq 0$. Let $z^*\equiv \left(E|z_{i,t}|^{p} \right)^{1/p}<\infty$, for  $p=\left\lbrace 1,2 \right\rbrace$ and the polynomial 
	\begin{equation}
		\phi(\lambda)=z^* \left( \beta_1 \lambda^{q+1} + \beta_2 \lambda^{q} + \cdots + \beta_q \lambda^{q-2}\right) - \lambda^{q+2}
		\label{eq:pol}
	\end{equation}
	has all roots $\lambda$ inside the unit circle. Then the process $r_{i,t}$ in (\ref{eq:larch_midas_x}) is weakly stationary.
	\label{theorem_1}
\end{theorem}
Proof: see Appendix \hyperref[sec:appendix_a]{A}. 

\subsection{Inference on the MF--QR--X Model \label{sec:estimation}}

In order to make inference on the MF--QR--X model, we need to solve Eq. (\ref{eq:check_fun_rho}) where
\begin{eqnarray}
{x}_{i-1,t}    &=&\left(1,|WS_{t-1}|,|r_{i-1,t}|,\ldots,|r_{i-q,t}|,|X_{i-t,t}|\right)' \label{eq:x_mf_q_arch_x} \\
{\Theta}_\tau &=&\left( \beta_{0,\tau},\theta_{\tau}, \beta_{1,\tau}, \ldots, \beta_ {q,\tau},  \beta_{X,\tau} \right).
\label{eq:Theta_mf_q_arch_x} 
\end{eqnarray}

The estimation of the vector ${\Theta}_\tau$ is encumbered by the fact that the mixed--frequency term $WS_{t-1}$ is not observable, as it depends on the unknown $\omega_2$ parameter of the weighting function $\delta_k(\omega)$, also to be estimated. To make estimation feasible, we resort to the expedient of profiling out\footnote{A profiling out strategy was used by \cite{Engle:Ghysels:Sohn:2013} for the parameter $K$ in the GARCH--MIDAS model.} the parameter $\omega_2$, through a two-step procedure: we first fix $\omega_2$ at an initial arbitrary value, say $\omega_2^{(b)}$, which turns the vector $x_{i-1,t}$ into a completely observable counterpart, in short ${x}_{i-1,t}^{(b)}$. This gives a solution to the minimization of the loss function, which is dependent on $\omega_2^{(b)}$, that is,
\begin{equation}
\widehat{\Theta}_\tau(\omega_2^{(b)}) \equiv \widehat{\Theta}^{(b)}_\tau = \argmin_{\Theta_\tau}  \sum  \rho_{\tau} \left( r_{i,t} - \left({x}_{i-1,t}^{(b)}\right)' {\Theta}_\tau \right).
\label{eq:prof_chech_loss}
\end{equation}
This procedure is repeated over a grid of $B$ values for $\omega_2$, so that we have $\left\lbrace\widehat{\Theta}^{(b)}_\tau\right\rbrace_{b=1}^B$, and the chosen overall estimator is  $\left(\hat\omega_2^*,\widehat{\Theta}^{(*)}_\tau\right)$, corresponding to the smallest overall value of the loss function.
	
Accordingly, the MF--QR--X estimator of the VaR is
\begin{equation}
\widehat{Q}_{r_{i,t}}\left(\tau|\mathcal{F}_{i-1,t}\right) = \left({x}_{i-1,t}^{(*)}\right)' \widehat{\Theta}^{(*)}_\tau.
	\label{eq:quant_est}
\end{equation} 

Summarizing, the proposed MF--QR--X is thus a flexible VaR model not requiring any distributional assumptions for the error term and accommodating both low--frequency and high-frequency additional variables. In Section \ref{sec:emp_analysis}, we will elaborate on its capability to jointly estimate the VaR and ES, adopting the approach proposed by \cite{Taylor:2019}. 

To obtain reliable VaR and ES estimates in our model (\ref{eq:quant_est}), an important issue is the choice of the optimal number of lags $q$  for the daily absolute returns in Eq. (\ref{eq:lin_arch}). To that end,  we select the lag order suggested by a sequential likelihood ratio ($LR$) test on individual lagged coefficients \citep[see also][]{Koenker:Machado:1999}. 
In particular, for a given $\tau$, at each step $j$ of the testing sequence over a range of $J$ values, we compare the unrestricted model where the number of lags is set equal to $j$ (labelled U, with an associated loss function ${V}_{U,\tau}^{(j)}$), against a restricted model where the number of lags is $j-1$ (labelled R, with an associated loss function ${V}_{R,\tau}^{(j-1)}$). In this setup, the null hypothesis of interest is 
\begin{equation}
H_0: \beta_j=0,
\label{eq:null_hyp}
\end{equation}
i.e., the coefficient on the most remote lag is zero.
The procedure starts contrasting a lag-1 model against a model with just a constant, then a lag-2 against a lag-1, and so on.  
For a given $\tau$, at each step $j$, we calculate the test statistic
 \begin{equation}
LR_{\tau}^{(j)}= \frac{2 \left({V}_{R,\tau}^{(j-1)}-{V}_{U,\tau}^{(j)} \right)}{\tau \left(1-\tau\right)s(\tau)},
\label{eq:LR}
\end{equation}
where  $s(\tau)$ is the so--called sparsity function estimated accordingly to \cite{Siddiqui:1960} and \cite{Koenker:Zhao:1996}. Under the adopted configuration, $LR_\tau^{(j)}$ is asymptotically distributed as a $\chi_1^{2}$, so that we select $q$ to be the last value of $j$ in the sequence, for which we reject the null hypothesis.

\section{Monte Carlo simulation \label{sec:monte_carlo}}

The finite sample properties of the sequential test and of the estimator of the MF--QR model\footnote{For simplicity, we have focused here on the case without the \virg{--X} component.} can be investigated by means of a Monte Carlo experiment. In what follows we consider $R=5000$ replications of the data generating process (DGP):
$$r_{i,t}=\left( \beta_0 + \theta |WS_{t-1}| + \beta_1|r_{i-1,t}| +\beta_2|r_{i-2,t}| +\beta_3|r_{i-3,t}| + \beta_4|r_{i-4,t}| \right)z_{i,t}, $$
where we assume a $\mathcal{N}(0,1)$ distribution for $z_{i,t}$ and we set to zero the relevant initial values for $r_{i,t}$. Moreover, the stationary variable $MV_t$ entering the weighted sum $WS_{t-1}$ is assumed to be drawn from an autoregressive AR(1) process $MV_t = \varphi MV_{t-1} + e_t$, with $\varphi=0.7$ and the error term $e_t$ following a Skewed $t$-distribution \citep{Hansen:1994}, with degrees of freedom $df=7$ and skewing parameter $sp=-6$. The frequency of $MV_{t}$ is monthly and $K=24$. The values of the parameters (collected in a vector $\Theta$) are detailed in the first column of the Tables \ref{tab:MC_0.01}--\ref{tab:MC_0.10}. For the simulation exercise we consider $N=1250$, $N=2500$ and $N=5000$ observations, to mimic realistic daily samples. \textcolor{black}{Having fixed $K=24$ (that is, two years of monthly data), the number of daily observations should be large enough to allows for model estimation. In our case, we set this limit to 1250 daily observations.}
Finally, three different levels of the VaR coverage level $\tau$ are chosen: 0.01, 0.05, and 0.10. 

In the Monte Carlo experiment, we start by evaluating the features of the $LR$ test for the lag selection in Eq. (\ref{eq:LR}). To that end, we test sequentially $H_0:\beta_j=0$ over $J$ steps at a significance level $\alpha$. Since the DGP is a fourth--order process, we expect to have a high rejection rate when the null involves a zero restriction on coefficients $\beta_j, \ j=1,\ldots,4$. In order to confirm the expected low rate of rejections, we extend the sequence of testing of further $\beta_j$'s, up to $J=6$.

Looking at the Table \ref{tab:seq_test}, where we report the percentages of rejections for different VaR coverage levels $\tau=0.01, 0.05, 0.1$ at the nominal significance level of $\alpha=5\%$ across replications, we validate the good behavior of the test.  Overall, the sequential test procedure satisfactorily identifies the number of lags to be included in the MF--QR model, with the performance improving with the number of observations, especially for $H_0:\beta_4=0$; for the latter case, the percentage of rejections of the null increases considerably across coverage levels when $N=5000$.

\begin{table}[htbp]
\caption{Percentage of rejection of the $LR$ test for the null $\beta_j=0$ \label{tab:seq_test}}       
\vspace{-0.2cm}   
\centering 
\begin{adjustbox}{max width=0.8\textwidth}
	\begin{threeparttable}
\begin{tabular}{l ... ... ...}
\toprule
& \multicolumn{9}{c}{VaR coverage}\\
    & \multicolumn{3}{c}{$\tau=0.01$}  & \multicolumn{3}{c}{$\tau=0.05$} & \multicolumn{3}{c}{$\tau=0.1$} \\
     \addlinespace
    \midrule   
       \addlinespace
 $N$   & \mc{1250} & \mc{2500} & \mc{5000}& \mc{1250} & \mc{2500} & \mc{5000}& \mc{1250} & \mc{2500} & \mc{5000}\\
 \addlinespace
 \hdashline
 \addlinespace
  
  $\beta_1=0$ 	&	99.56	&	100.00	&	100.00	&	100.00	&	100.00	&	100.00	&	100.00	&	100.00	&	100.00	\\
  $\beta_2=0$ 	&	97.66	&	100.00	&	100.00	&	99.88	&	100.00	&	100.00	&	99.82	&	100.00	&	100.00	\\
  $\beta_3=0$ 	&	85.76	&	99.14	&	100.00	&	95.60	&	99.94	&	100.00	&	94.34	&	99.86	&	100.00	\\
  $\beta_4=0$ 	&	52.88	&	82.84	&	98.44	&	69.36	&	93.16	&	99.70	&	65.78	&	90.70	&	99.58	\\
  $\beta_5=0$ 	&	4.46	&	4.98	&	4.90	&	5.38	&	6.30	&	6.56	&	6.40	&	6.34	&	6.56	\\
  $\beta_6=0$ 	&	4.86	&	5.24	&	5.00	&	5.14	&	5.56	&	5.80	&	5.16	&	5.58	&	5.92	\\
%
\bottomrule
\addlinespace
\end{tabular}
\vspace{-0.3cm}
	\begin{tablenotes}[flushleft]
   \setlength\labelsep{0pt}
   \linespread{1.25}
\item  \textbf{Notes}: The table presents the percentage of rejection for the null in the first column, across all the Monte Carlo replicates, for three different
configurations of $N$ and $\tau$.
\end{tablenotes}
\end{threeparttable}
\end{adjustbox}
\end{table}

Turning to the small sample properties of our estimator, the evaluation is done in terms of the original coefficients in the DGP, collected in the vector $\Theta= \left(\beta_0,\theta,\beta_1,\ldots,\beta_q\right)$, using the relationship with the quantile regression parameters $\Theta_\tau$, i.e., $\Theta=\Theta_\tau / F^{-1}(\tau)$.  
\footnote{As per the parameter $\omega_2$, the grid search is done over 100 values and the applied rescaling factor is equal to $1$, as its value is unaffected by $\tau$.} In Tables \ref{tab:MC_0.01}, \ref{tab:MC_0.05} and \ref{tab:MC_0.10} we report the Monte Carlo averages of the parameters ($\hat\Theta$) across replications for three levels of $\tau$, and the estimated Mean Squared Errors relative to the true values. 

Overall, the proposed model presents good finite sample properties: independently of the $\tau$ level chosen, for small sample sizes, the estimates appear, in general, slightly biased, although, reassuringly, the MSE of the estimates relative to the true values always decreases as the sample period increases.

\begin{table}[htbp]
\centering
\caption{Monte Carlo estimates, $\tau=0.01$ \label{tab:MC_0.01}}       
\vspace{-0.2cm}
\centering
	\begin{adjustbox}{max width=0.7\textwidth}
	\begin{threeparttable}
	\begin{tabular}{l . ........}
\toprule
 &     \mc{True $\Theta$} &    \mc{\hspace{3mm} $\overline{\widehat\Theta}$} &    \mc{\hspace{3mm} MSE}    &       \mc{\hspace{3mm} $\overline{\widehat\Theta}$} &    \mc{\hspace{3mm} MSE}  & \mc{\hspace{3mm} $\overline{\widehat\Theta}$} &    \mc{\hspace{3mm} MSE} \\
\midrule
&&\multicolumn{2}{c}{\hspace{3mm}$N=1250$}&\multicolumn{2}{c}{\hspace{3mm}$N=2500$}&\multicolumn{2}{c}{\hspace{3mm}$N=5000$}\\
$\beta_0$ & 0.050 & 0.079 & 0.040 & 0.064 & 0.019 & 0.058 & 0.009 \\ 
  $\theta$ & 0.125 & 0.124 & 0.013 & 0.126 & 0.007 & 0.125 & 0.003 \\ 
  $\beta_1$ & 0.300 & 0.286 & 0.009 & 0.292 & 0.005 & 0.296 & 0.002 \\ 
  $\beta_2$ & 0.250 & 0.236 & 0.008 & 0.242 & 0.004 & 0.246 & 0.002 \\ 
  $\beta_3$ & 0.200 & 0.187 & 0.008 & 0.194 & 0.004 & 0.196 & 0.002 \\ 
  $\beta_4$ & 0.150 & 0.143 & 0.007 & 0.146 & 0.004 & 0.149 & 0.002 \\ 
   \addlinespace
 \hdashline
 \addlinespace
  $\omega_2$ & 2.000 & 1.993 & 0.010 & 1.991 & 0.010 & 1.984 & 0.010 \\ 
\bottomrule
 \end{tabular}
	\begin{tablenotes}[flushleft]
   \setlength\labelsep{0pt}
   \linespread{1.25}
\item  \textbf{Notes}: The first column shows the true values of the $\Theta$ coefficients in the DGP. Simulations were replicated 5000 times, according to three different window lengths: $N=1250$, $N=2500$, and $N=5000$. Columns $\overline{\widehat\Theta}$ report the averages of the estimated parameters across replications. Columns labeled \textit{MSE} refer to the Mean Square Error of the estimated coefficients relative to the true values.
\end{tablenotes}
\end{threeparttable}
\end{adjustbox}
\end{table}

\begin{table}[htbp]
\centering
\caption{Monte Carlo estimates, $\tau=0.05$ \label{tab:MC_0.05}}       
\vspace{-0.2cm}
\centering
	\begin{adjustbox}{max width=0.7\textwidth}
	\begin{threeparttable}
	\begin{tabular}{l . ........}
\toprule
 &     \mc{True $\Theta$} &    \mc{\hspace{3mm} $\overline{\widehat\Theta}$} &    \mc{\hspace{3mm} MSE}    &       \mc{\hspace{3mm} $\overline{\widehat\Theta}$} &    \mc{\hspace{3mm} MSE}  & \mc{\hspace{3mm} $\overline{\widehat\Theta}$} &    \mc{\hspace{3mm} MSE} \\
\midrule
&&\multicolumn{2}{c}{\hspace{3mm}$N=1250$}&\multicolumn{2}{c}{\hspace{3mm}$N=2500$}&\multicolumn{2}{c}{\hspace{3mm}$N=5000$}\\
$\beta_0$ & 0.050 & 0.066 & 0.025 & 0.057 & 0.012 & 0.053 & 0.006 \\ 
  $\theta$ & 0.125 & 0.123 & 0.008 & 0.125 & 0.004 & 0.125 & 0.002 \\ 
  $\beta_1$ & 0.300 & 0.294 & 0.006 & 0.297 & 0.003 & 0.299 & 0.002 \\ 
  $\beta_2$ & 0.250 & 0.242 & 0.006 & 0.246 & 0.003 & 0.248 & 0.001 \\ 
  $\beta_3$ & 0.200 & 0.195 & 0.005 & 0.196 & 0.003 & 0.198 & 0.001 \\ 
  $\beta_4$ & 0.150 & 0.146 & 0.005 & 0.148 & 0.002 & 0.149 & 0.001 \\ 
 \addlinespace
 \hdashline
 \addlinespace
   $\omega_2$ & 2.000 & 1.991 & 0.010 & 1.985 & 0.010 & 1.977 & 0.009 \\ 
\bottomrule
 \end{tabular}
	\begin{tablenotes}[flushleft]
   \setlength\labelsep{0pt}
   \linespread{1.25}
\item  \textbf{Notes}: The first column shows the true values of the $\Theta$ coefficients in the DGP. Simulations were replicated 5000 times, according to three different window lengths: $N=1250$, $N=2500$, and $N=5000$. Columns $\overline{\widehat\Theta}$ report the averages of the estimated parameters across replications. Columns labeled \textit{MSE} refer to the Mean Square Error of the estimated coefficients relative to the true values.
\end{tablenotes}
\end{threeparttable}
\end{adjustbox}
\end{table}

\begin{table}[htbp]
\centering
\caption{Monte Carlo estimates, $\tau=0.1$ \label{tab:MC_0.10}}       
\vspace{-0.2cm}
\centering
	\begin{adjustbox}{max width=0.7\textwidth}
	\begin{threeparttable}
	\begin{tabular}{l . ........}
\toprule
 &     \mc{True $\Theta$} &    \mc{\hspace{3mm} $\overline{\widehat\Theta}$} &    \mc{\hspace{3mm} MSE}    &       \mc{\hspace{3mm} $\overline{\widehat\Theta}$} &    \mc{\hspace{3mm} MSE}  & \mc{\hspace{3mm} $\overline{\widehat\Theta}$} &    \mc{\hspace{3mm} MSE} \\
\midrule
&&\multicolumn{2}{c}{\hspace{3mm}$N=1250$}&\multicolumn{2}{c}{\hspace{3mm}$N=2500$}&\multicolumn{2}{c}{\hspace{3mm}$N=5000$}\\
$\beta_0$ & 0.050 & 0.063 & 0.026 & 0.057 & 0.013 & 0.053 & 0.006 \\ 
  $\theta$ & 0.125 & 0.124 & 0.009 & 0.124 & 0.004 & 0.125 & 0.002 \\ 
  $\beta_1$ & 0.300 & 0.296 & 0.006 & 0.297 & 0.003 & 0.299 & 0.002 \\ 
  $\beta_2$ & 0.250 & 0.244 & 0.006 & 0.246 & 0.003 & 0.248 & 0.002 \\ 
  $\beta_3$ & 0.200 & 0.196 & 0.006 & 0.198 & 0.003 & 0.199 & 0.002 \\ 
  $\beta_4$ & 0.150 & 0.145 & 0.005 & 0.148 & 0.003 & 0.149 & 0.001 \\ 
   \addlinespace
 \hdashline
 \addlinespace
   $\omega_2$ & 2.000 & 1.992 & 0.010 & 1.986 & 0.010 & 1.979 & 0.009 \\
\bottomrule
 \end{tabular}
	\begin{tablenotes}[flushleft]
   \setlength\labelsep{0pt}
   \linespread{1.25}
\item  \textbf{Notes}: The first column shows the true values of the $\Theta$ coefficients in the DGP. Simulations were replicated 5000 times, according to three different window lengths: $N=1250$, $N=2500$, and $N=5000$. Columns $\overline{\widehat\Theta}$ report the averages of the estimated parameters across replications. Columns labeled \textit{MSE} refer to the Mean Square Error of the estimated coefficients relative to the true values.
\end{tablenotes}
\end{threeparttable}
\end{adjustbox}
\end{table}

\section{Model Evaluation \label{sec:back_model_selec}}

In order to evaluate the quality of the tail risk estimates we can resort to a set of tests suitable to the needs of risk management. Above all, the backtesting procedure is very popular in evaluating risk measure performance \citep[see the reviews of][among others]{Campbell:2006, Nieto:Ruiz:2016}. For our model we use the Actual over Expected (AE) exceedance ratio and five other tests in this class: the Unconditional Coverage \citep[UC,][]{Kupiec:1995}, the Conditional
Coverage \citep[CC,][]{Christoffersen:1998}, and the Dynamic Quantile \citep[DQ,][]{Engle:Manganelli:2004} tests for the VaR and the UC and CC tests for the ES \citep{Acerbi:Szekely:2014}. 

The AE exceedance ratio is the number of times that the VaR measures have been violated over the expected VaR violations. The closer to one the ratio, the better is the model to forecast VaRs. The UC test is a $LR$--based test, where the null hypothesis assesses whether the actual frequency of VaR violations is equal to the chosen $\tau$ level. \textcolor{black}{Formally, the null hypothesis of the UC test is 
$$H_0:  \pi = \tau,$$ 
where $\pi=\mathbb{E}[L_{i,t}(\tau)]$, with $L_{i,t}(\tau)=\mathbbm{1}_{\left(r_{i,t}<VaR_{i,t}(\tau)\right)}$ representing the series of VaR violations.} The UC test statistic is asymptotically $\chi^2$ distributed, with one degree of freedom, \textcolor{black}{assuming independence of the $L_{i,t}(\tau)$ series}. 

\textcolor{black}{Another critical aspect to test for is the independence of VaR violations over time. The main idea is to discard models whose VaR forecasts are violated in subsequent days. Moreover, if the assumption of independence is not satisfied by the
violations, the asymptotic results on the distribution of the UC test can fail to hold. The independence test used in this context is that of \cite{Christoffersen:1998}, where the null hypothesis consists of independence of $L_{i,t}(\tau)$, while the alternative hypothesis is that $L_{i,t}(\tau)$ follows a first-order Markov Chain. Under $H_0$, the $LR$--based test is asymptotically $\chi^2$ distributed, with one degree of freedom.} 

\textcolor{black}{An overall assessment of the VaR measures is given by the CC test conducted on both null hypotheses of the UC and of the independence tests jointly (asymptotically the test statistic is $\chi^2$ distributed, with two degrees of freedom)}. 

The DQ test also applies to the independence of the VaR violations jointly with the correctness of the number of violations as the CC test, but it was shown \citep{Berkowitz:Christoffersen:Pelletier:2011} to have more power over it. In particular, the DQ test consists of running a linear regression where the dependent variable is the sequence of VaR violations and the covariates are the past violations and possibly any other explanatory variables.  \textcolor{black}{More in detail, let $Hit_{i,t}(\tau)=L_{i,t}(\tau)-\tau$ be the so-called series of the \emph{hit} variable. This series, under correct specification, should have zero mean, be serially uncorrelated and, moreover, uncorrelated with any other past observed variables. The DQ test can be carried via the following OLS regression:
\begin{equation}
Hit_{i,t}(\tau)=\beta_0+\sum_{k=1}^{K_1} \beta_k \, Hit_{i-k,t}(\tau) + \sum_{k=1}^{K_2} \gamma_k \, Z_{i-k,t}(\tau) + u_{i,t}, \label{eq:ols_reg}
\end{equation}
where $u_{i,t}$ is the error term and $Z_{i,t}(\tau)$'s include potentially relevant variables belonging to the available information set, like, for instance, previous $Hits$, lagged VaR or past returns. In matrix notation, the  OLS regression in \eqref{eq:ols_reg} becomes:
\begin{equation}
\bm{Hit}=\bm{Z}\bm{\psi} + \bm{u},
\label{eq:dq_mat}
\end{equation}
where the vector $\bm{Hit}$ has dimension $N$ (with $N$ indicating the total number of observations),  the matrix of predictors $\bm{Z}$ has dimension $N \times (K_1+K_2+1)$, the vector $\bm{\psi}=\left(\beta_0,\beta_1,\cdots,\beta_{K_1},\gamma_1,\cdots,\gamma_{K_2} \right)$ has dimension $(K_1+K_2+1)$, and the error vector $\bm{u}$ has dimension $N$. Under correct specification we test  the null $\bm{\psi}=\bm{0}$ with a test statistic:
$$DQ_{CC}=\frac{\hat{\bm{\psi}}^{'}\bm{Z}^{'}\bm{Z}\hat{\bm{\psi}}}{\tau(1-\tau)}\overset{d}{\rightarrow}\chi_{K_1+K_2+1}^2,$$
where $\hat{\bm{\psi}}$ is the estimated vector of coefficients obtained from the OLS regression in \eqref{eq:dq_mat}.}

\textcolor{black}{For the expected shortfall $ES$, the UC test of \cite{Acerbi:Szekely:2014} is based on the following statistic:
\begin{equation}
Z_{UC}=\frac{1}{N(1-\tau)}\sum_{i=1}^{N_t}\sum_{t=1}^{T} \frac{r_{i,t}L_{i,t}(\tau)}{ES_{i,t}(\tau)}+1.
\end{equation}
If the distributional assumptions are correct, the expected value of $Z_{UC}$ is zero, that is $\mathbb{E}\left(Z_{UC}\right)=0$. The CC test of \cite{Acerbi:Szekely:2014} has the following statistic:
\begin{equation}
Z_{CC}=\frac{1}{NumFail}\sum_{i=1}^{N_t}\sum_{t=1}^{T} \frac{r_{i,t}L_{i,t}(\tau)}{ES_{i,t}(\tau)}+1,
\end{equation}
where $NumFail=\sum_{i=1}^{N_t}\sum_{t=1}^{T}L_{i,t}(\tau)$. If the distributional assumptions are correct, the expected value of $Z_{CC}$, given that there is at least one $VaR$ violation, is zero, i.e. $\mathbb{E}\left(Z_{CC}|NumFail>0\right)=0$. The UC and CC tests are one-sided and reject the null when the model underestimates the risk (significantly negative test statistic).}

\section{Empirical Analysis \label{sec:emp_analysis}}

\begin{figure}[htbp]
\caption{Crude Oil, Gasoline, VIX and GPR}
\label{fig:returns}       
\vspace{-0.2cm}
\includegraphics[width=1\textwidth]{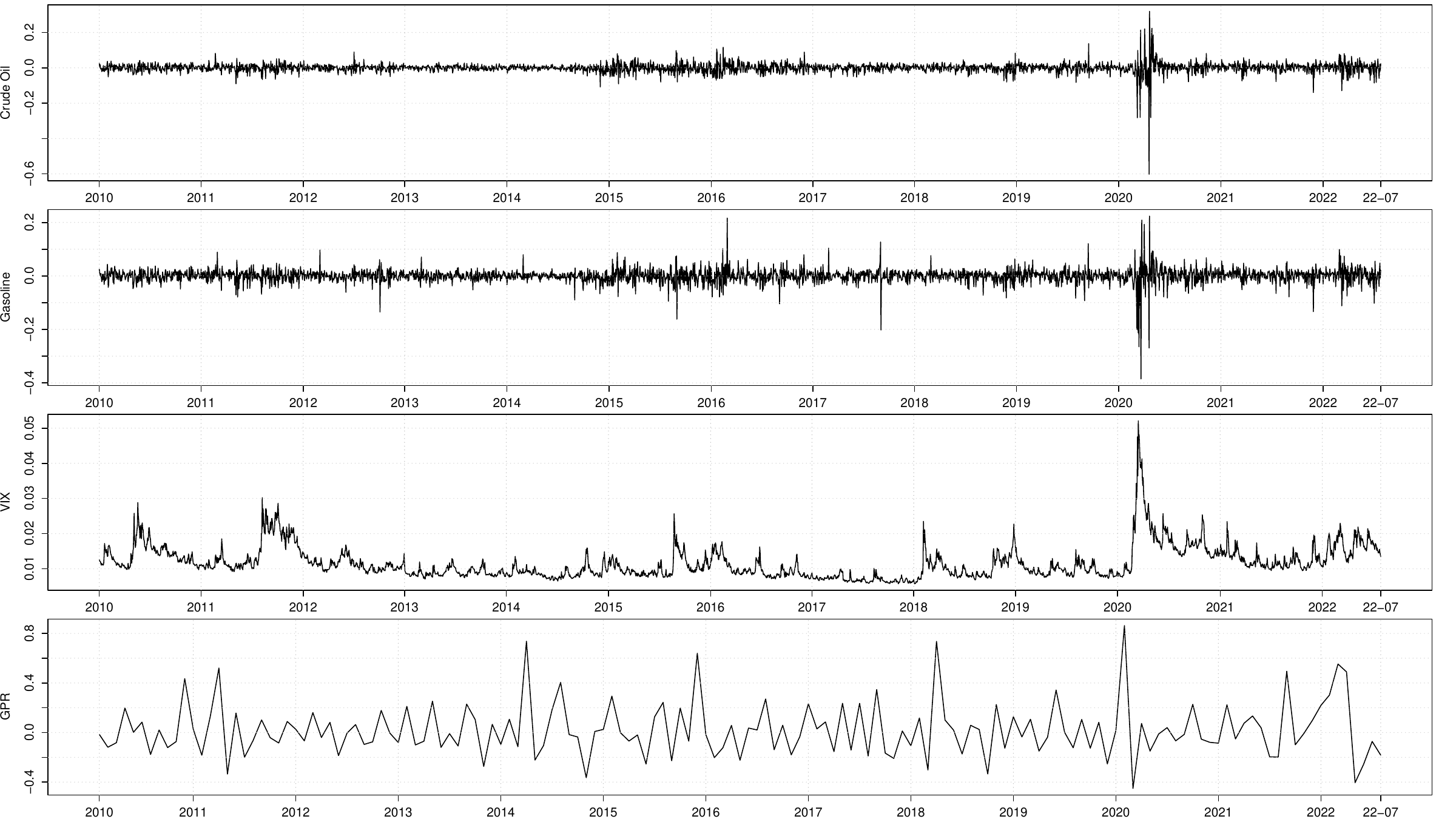}
\end{figure}

In this section, we apply the MF--QR--X model to estimate\footnote{\textcolor{black}{In terms of computational efforts, it is worth noting that the proposed MF--QR--X model is not excessively demanding. For instance, VaR and ES (via maximization of the ALD) are obtained in 27 seconds, considering five years of data, with the --X variable, on a standard PC (AMD A10-9700 RADEON R7, 10 COMPUTE CORES 4C+6G, 3.50 GHz, 12 GB of RAM).}} VaR and ES for the daily log-returns of two energy commodities: the WTI Crude Oil and the RBOB Gasoline futures.\footnote{Both the WTI and RBOB futures have been downloaded from the Yahoo Finance site (with, respectively, ticks \virg{CL=F} and \virg{RBOB=F}).} The low--frequency variable is the monthly GPR index, which enters our mixed--frequency models as the first difference divided by one lagged realization. The \virg{--X} variable is the VIX index.\footnote{Taken from the Yahoo finance site and transformed by dividing it by $\sqrt{252} \cdot 100$, in order to express it as daily volatility.} The period of investigation covers almost 13 years, from January 2010 to July 2022 on a daily basis, split between  in-- (from January 2010 to December 2016) and out--of--sample periods (from January 2017 to July 2022).  The data are summarized in Table \ref{tab:sum_stat_1}, and plotted in Figure \ref{fig:returns}. 

\begin{table}[htbp]
	\centering
		\caption{Summary statistics}
		\vspace{-0.2cm}
	\label{tab:sum_stat_1}
\begin{adjustbox}{max width=0.75\textwidth}
	\begin{threeparttable}
\begin{tabular}{l  .......}
\toprule
 &     \mc{Obs.} &          \mc{Min.} &         \mc{Max.} &      \mc{Mean} &     \mc{SD} &       \mc{Skew.} & \mc{Kurt.} \\
\midrule
&\multicolumn{7}{c}{Full sample: 2010/2022-07}\\
 \addlinespace
Crude Oil & 3160 & -0.602 & 0.320 & 0.006 & 0.029 & -2.840 & 80.150 \\ 
  Gasoline & 3159 & -0.385 & 0.224 & 0.016 & 0.027 & -1.603 & 30.230 \\ 
  VIX & 3159 & 0.006 & 0.052 & 1.165 & 0.005 & 2.359 & 10.367 \\ 
  GPR & 151 & -0.451 & 0.863 & 2.250 & 0.219 & 1.093 & 2.045 \\
               \addlinespace
        \hdashline
        \addlinespace
 &\multicolumn{7}{c}{In-sample: 2010/2016}\\      
  \addlinespace 
 Crude Oil & 1760 & -0.108 & 0.116 & -0.022 & 0.021 & 0.131 & 2.877 \\ 
  Gasoline & 1759 & -0.162 & 0.217 & -0.012 & 0.022 & 0.131 & 9.388 \\ 
  VIX & 1759 & 0.007 & 0.030 & 1.129 & 0.004 & 1.760 & 3.506 \\ 
  GPR & 84 & -0.364 & 0.737 & 1.851 & 0.197 & 1.099 & 1.956 \\
  \addlinespace
        \hdashline
        \addlinespace
 &\multicolumn{7}{c}{Out-of-sample: 2017/2022-07}\\ 
  \addlinespace
 Crude Oil & 1400 & -0.602 & 0.320 & 0.042 & 0.036 & -3.334 & 73.037 \\ 
  Gasoline & 1400 & -0.385 & 0.224 & 0.050 & 0.031 & -2.268 & 31.855 \\ 
  VIX & 1400 & 0.006 & 0.052 & 1.210 & 0.006 & 2.300 & 9.368 \\ 
  GPR & 67 & -0.451 & 0.863 & 2.750 & 0.244 & 1.020 & 1.621 \\
 \bottomrule
\addlinespace
 \end{tabular}
\vspace{-0.2cm}
	\begin{tablenotes}[flushleft]
   \setlength\labelsep{0pt}
   \linespread{1.25}
\item  \textbf{Notes}:  The table reports the number of observations (Obs.), the minimum (Min.) and maximum (Max.), the mean (multiplied by 100), the standard deviation (SD), the Skewness (Skew.) and the excess kurtosis (Kurt.).  The variables are: the daily close-to-close log-returns of WTI Crude Oil and RBOB Gasoline, the daily VIX and the first difference of the monthly GPR index divided by its lagged realization.
\end{tablenotes}
\end{threeparttable}
	\end{adjustbox} 
\end{table}

We compare the estimated VaR and ES with several well--known competitive specifications belonging to the class of parametric  (GARCH, GJR \citep{Glosten:Jaganathan:Runkle:1993}, and GARCH--MIDAS, with Gaussian and Student's t error distributions), non-parametric (HS) and semi-parametric models (the Symmetric Absolute Value (SAV), Asymmetric Slope (AS) and Indirect GARCH (IG) specifications of the CAViaR \citep{Engle:Manganelli:2004}). As per the mixed--frequency specifications, the same low--frequency variable (GPR index) is inserted as the low--frequency variable in the GARCH-MIDAS specifications as well as our proposed MF--QR and MF--QR--X models. All the functional forms of these models are reported in Table \ref{tab:models_eq}. 

\begin{table}[htbp]
  \centering
  \caption{Model specifications \label{tab:models_eq}}
  \vspace{-0.2cm}
  \begin{adjustbox}{max width=0.8\textwidth}
    \begin{threeparttable}
      \begin{tabular}{l c c}
        \toprule  
        Model      &   Functional form     & Err. Distr.\\
        \midrule                                                                                                          
            \multirow{2}{*}{GARCH--N}              & $r_{i,t}|\mathcal{F}_{i-1,t}  = \sqrt{h_{i,t}} \eta_{i,t}$ & $\eta_{i,t}\overset{i.i.d}{\sim} \mathcal{N}\left(0, 1\right)$\\  
           &      $h_{i,t} = \omega + \alpha_{}  r_{i-1, t}^2 + \beta h_{i-1, t}$& \\
        \addlinespace
        \hdashline
        \addlinespace
           \multirow{2}{*}{GARCH--t}              & $r_{i,t}|\mathcal{F}_{i-1,t}  = \sqrt{h_{i,t}} \eta_{i,t}$ & $\eta_{i,t}\overset{i.i.d}{\sim} t_{\nu}$\\  
           &      $h_{i,t} = \omega + \alpha  r_{i-1, t}^2 + \beta h_{i-1, t}$& \\     
               \addlinespace
        \hdashline
        \addlinespace
         \multirow{2}{*}{GJR--N}              & $r_{i,t}|\mathcal{F}_{i-1,t}  = \sqrt{h_{i,t}} \eta_{i,t}$ & $\eta_{i,t}\overset{i.i.d}{\sim} \mathcal{N}\left(0, 1\right)$\\  
           &      $h_{i,t} = \omega + \left(\alpha_{} + \gamma_{} \mathbbm{1}_{\left( r_{i-1, t} < 0 \right)} \right) r_{i-1, t}^2 + \beta_{} h_{i-1, t}$& \\
          \addlinespace
        \hdashline
        \addlinespace  
                \multirow{2}{*}{GJR--t}              & $r_{i,t}|\mathcal{F}_{i-1,t}  = \sqrt{h_{i,t}} \eta_{i,t}$ & $\eta_{i,t}\overset{i.i.d}{\sim} t_{\nu}$\\  
           &      $h_{i,t} = \omega + \left(\alpha_{} + \gamma_{} \mathbbm{1}_{\left( r_{i-1, t} < 0 \right)} \right) r_{i-1, t}^2 + \beta_{} h_{i-1, t}$& \\
                     \addlinespace
        \hdashline
        \addlinespace  
    
                & $r_{i,t}|\mathcal{F}_{i-1,t}  = \sqrt{\pi_t \times \xi_{i,t} } \eta_{i,t}$ & $\eta_{i,t}\overset{i.i.d}{\sim} \mathcal{N}\left(0, 1\right)$\\    
        \textbf{GM--N}  &      $\xi_{i,t}=(1-\alpha_{}-\beta_{}-\gamma_{}/2) + \left(\alpha_{} +  \gamma_{} \cdot \mathbbm{1}_{\left(r_{i-1,t}  < 0 \right)}\right) \frac{r_{i-1,t}^2}{\pi_t} + \beta_{} \xi_{i-1,t}$& \\
        & $\pi_t=\exp\left\lbrace m + \zeta \sum_{k=1}^K \delta_k(\omega) MV_{t-k}\right\rbrace$\\
                         \addlinespace
        \hdashline
        \addlinespace  
    
                & $r_{i,t}|\mathcal{F}_{i-1,t}  = \sqrt{\pi_t \times \xi_{i,t} } \eta_{i,t}$ & $\eta_{i,t}\overset{i.i.d}{\sim} t_{\nu}$\\    
        \textbf{GM--t}  &      $\xi_{i,t}=(1-\alpha_{}-\beta_{}-\gamma_{}/2) + \left(\alpha_{} +  \gamma_{} \cdot \mathbbm{1}_{\left(r_{i-1,t}  < 0 \right)}\right) \frac{r_{i-1,t}^2}{\pi_t} + \beta_{} \xi_{i-1,t}$& \\
        & $\pi_t=\exp\left\lbrace m + \zeta \sum_{k=1}^K \delta_k(\omega) MV_{t-k}\right\rbrace$\\
           \addlinespace
        \hdashline
        \addlinespace        
        
  \multirow{2}{*}{HS} & $VaR_{i,t}(\tau)={Q}_{\bm r_{i,t}^{w}}(\tau)$\\ 
  &$\bm r_{i,t}^{w} = (r_{i-w,t},r_{i-w+1,t},\dots,r_{i-1,t})$    \\ 
        
  \addlinespace
        \hdashline
        \addlinespace    
      
  SAV       &     $VaR_{i,t}(\tau) = \beta_0+ \beta_1 VaR_{i-1,t}(\tau)+ \beta_2 |r_{i-1,t}| $&\\

        \addlinespace
        \hdashline
        \addlinespace
          AS       &    $VaR_{i,t}(\tau) =  \beta_0+  \beta_1 VaR_{i-1,t}(\tau)+ (\beta_{2}\mathbbm{1}_{(r_{i-1,t}>0)}+\beta_{3} \mathbbm{1}_{(r_{i-1,t}<0)}) |r_{i-1,t}|$&\\
        \addlinespace
        \hdashline
        \addlinespace
         IG      &    $VaR_{i,t}(\tau) =  -\sqrt {\beta_0+ \beta_1 VaR_{i-1,t}^2(\tau)+ \beta_2 r_{i-1,t}^2 }$&\\ 
           \addlinespace
        \hdashline
        \addlinespace
    
 \multirow{2}{*}{QR}       & $r_{i,t}|\mathcal{F}_{i-1,t}  =\sigma_{i,t}  z_{i,t}$ & $z_{i,t}\overset{i.i.d}{\sim} \left(0, 1\right)$\\         
         &   $\sigma_{i,t}=  \left(\beta_0 + \beta_1 |r_{i-1,t}| + \ldots + \beta_q |r_{i-q,t}|   \right)$ &\\
                         \addlinespace
        \hdashline
        \addlinespace  
        \multirow{2}{*}{QR--X}       & $r_{i,t}|\mathcal{F}_{i-1,t}  =\sigma_{i,t}  z_{i,t}$ & $z_{i,t}\overset{i.i.d}{\sim} \left(0, 1\right)$\\         
         &   $\sigma_{i,t}=  \left(\beta_0 + \beta_1 |r_{i-1,t}| + \ldots + \beta_q |r_{i-q,t}| +  \beta_X |X_{i-1,t}|  \right)$ &\\
                                  \addlinespace
        \hdashline
        \addlinespace  
          & $r_{i,t}|\mathcal{F}_{i-1,t}  =\sigma_{i,t}  z_{i,t}$ & $z_{i,t}\overset{i.i.d}{\sim} \left(0, 1\right)$\\         
  \textbf{MF--QR}        &   $\sigma_{i,t}=  \left(\beta_0 + \theta |WS_{t-1}|+ \beta_1 |r_{i-1,t}| + \ldots + \beta_q |r_{i-q,t}|   \right)$ &\\
        & $WS_{t-1}= \sum_{k=1}^K \delta_k(\omega)MV_{t-k}$&\\
                                        \addlinespace
        \hdashline
        \addlinespace 
        & $r_{i,t}|\mathcal{F}_{i-1,t}  =\sigma_{i,t}  z_{i,t}$ & $z_{i,t}\overset{i.i.d}{\sim} \left(0, 1\right)$\\         
  \textbf{MF--QR--X}        &   $\sigma_{i,t}=  \left(\beta_0 + \theta |WS_{t-1}|+ \beta_1 |r_{i-1,t}| + \ldots + \beta_q |r_{i-q,t}| + \beta_X |X_{i-1,t}|  \right)$ &\\
        & $WS_{t-1}= \sum_{k=1}^K \delta_k(\omega)MV_{t-k}$&\\

        \bottomrule
      \end{tabular}
      \begin{tablenotes}[flushleft]
        \setlength\labelsep{0pt}
        \footnotesize
        \item \textbf{Notes}: The table reports the functional forms for the parametric models, that is GARCH, GJR, and GARCH-MIDAS models, with Gaussian and Student's t distributions for the errors (GARCH-N, GARCH-t, GJR-N, GJR-t, GM-N, and GM-t, respectively), the non-parametric models (Historical Simulations with length window $w$, \citep[HS,][]{Hendricks:1996}) and the semi-parametric models, that is Aymmetric Absolute Value (SAV), Asymmetric Slope (AS), Indirect GARCH (IG), Quantile regression \citep[QR,][]{Koenker:Zhao:1996}, Quantile Regression with X component (QR--X), Mixed--Frequency Quantile Regression (MF--QR) and Mixed--Frequency Quantile Regression with X component (MF--QR--X) models. Labels in bold indicate models using a low--frequency variable.
      \end{tablenotes}
    \end{threeparttable}
  \end{adjustbox}
\end{table}

\subsection*{In-sample analysis}

Tables \ref{tab:seq_test_emp_analysis} reports the p-values of the  $LR$ test (Equation \ref{eq:LR}) using $\tau=0.05$, on the period from  2010 to 2016, for the two commodities under investigation, which suggests the inclusion of up to six, respectively, five lagged daily log--returns in the models for the Crude Oil and Gasoline futures. 

\begin{table}[h]
	\centering
		\caption{$LR$ test, p-values of the null $\beta_j=0$ 	\label{tab:seq_test_emp_analysis}}
\vspace{-0.2cm}
\centering
	\begin{adjustbox}{max width=0.75\textwidth}
	\begin{threeparttable}
		\begin{tabular}{lccccccccc}
\toprule
 Index & $\beta_1=0$& $\beta_2=0$ &$\beta_3=0$& $\beta_4=0$ &$\beta_5=0$ & $\beta_6=0$& $\beta_7=0$ & $\beta_8=0$&$\beta_9=0$ \\ 
\midrule
Crude Oil & 0.000 &  0.000 &   0.037  &  0.017   & 0.146   & 0.002  &  0.059  &  0.464  &  0.514\\ 
Gasoline & 0.000 & 0.022  &  0.063 &    0.820  &  0.012  &   0.370    &0.242 &   0.139   & 0.984\\ 
 \bottomrule
 \end{tabular}
	\begin{tablenotes}[flushleft]
   \setlength\labelsep{0pt}
   \linespread{1.25}
\item  \textbf{Notes}:  The table reports the p-values of $LR$ test according to the procedure highlighted in Section \ref{sec:estimation}, for the null in column. Sample period: from January 2010 to December 2016.
\end{tablenotes}
\end{threeparttable}
	\end{adjustbox}
\end{table}

As regards the number of lagged realizations entering the low--frequency component, we choose $K=36$, for all mixed frequency models. The in-sample estimated parameters for the parametric  \citep[with Quasi Maximum Likelihood  standard errors, cf.][]{Bollerslev:Wooldridge:1992} and semi-parametric models \citep[with bootstrap-based standard errors, as done also by][]{Xu:Wang:Liu:2021} are reported in Tables \ref{tab:in_sample_est_crude_oil} (Crude Oil) and \ref{tab:in_sample_est_gasoline} (Gasoline). \textcolor{black}{The algorithm used to obtain the bootstrap standard errors is sketched in Appendix \hyperref[sec:appendix_b]{B}}. Note that for the proposed MF--QR--X model, the low--frequency parameters as well as the parameters associated to the \virg{--X} variable are generally significant.

\begin{landscape}
\begin{table}
    \centering
\caption{In-sample estimates for Crude Oil \label{tab:in_sample_est_crude_oil}}
  \vspace{-0.2cm}
\begin{adjustbox}{max height=0.22\textheight}
\begin{threeparttable}
\begin{tabular}{l  ................}
\toprule
	&	        \mc{$\omega$} 	&	       \alpha 	&	       \beta 	&	     \gamma 	&	            m 	&	      \theta 	&	    \omega_2 	&	          \nu 	&	       \beta_1 	&	      \beta_2 	&	       \beta_3 	&	       \beta_4 	&	       \beta_5 	&	       \beta_6  &	       \beta_X	&	     \gamma_{ES}	\\
\midrule
GARCH--N	&	        0.000^{ } 	&	  0.074^{ } 	&	 0.922^{***} 	&	      	&	         	&	         	&	      	&	        	&	         	&	         	&	         	&	       	&	       	&	         	&	         	&	         	\\
	&	          (0.000) 	&	    (0.182) 	&	     (0.203) 	&	      	&	         	&	         	&	      	&	        	&	         	&	         	&	         	&	       	&	       	&	         	&	         	&	         	\\
GARCH--t  	&	        0.000^{ } 	&	   0.07^{ } 	&	 0.926^{***} 	&	      	&	         	&	         	&	      	&	  7.99^{***} 	&	         	&	         	&	         	&	       	&	       	&	         	&	         	&	         	\\
	&	          (0.000) 	&	    (0.086) 	&	     (0.092) 	&	      	&	         	&	         	&	      	&	     (1.751) 	&	         	&	         	&	         	&	       	&	       	&	         	&	         	&	         	\\
GJR--N	&	        0.000^{ } 	&	  0.011^{ } 	&	   0.939^{ } 	&	 0.091^{ } 	&	         	&	         	&	      	&	        	&	         	&	         	&	         	&	       	&	       	&	         	&	         	&	         	\\
	&	          (0.000) 	&	    (0.292) 	&	     (0.673) 	&	   (0.673) 	&	         	&	         	&	      	&	        	&	         	&	         	&	         	&	       	&	       	&	         	&	         	&	         	\\
GJR--t	&	        0.000^{ } 	&	  0.011^{ } 	&	 0.946^{***} 	&	 0.078^{ } 	&	         	&	         	&	      	&	 8.931^{***} 	&	         	&	         	&	         	&	       	&	       	&	         	&	         	&	         	\\
	&	          (0.000) 	&	    (0.031) 	&	     (0.056) 	&	   (0.048) 	&	         	&	         	&	      	&	     (2.341) 	&	         	&	         	&	         	&	       	&	       	&	         	&	         	&	         	\\
GM--N	&	         	&	 0.071^{**} 	&	  0.92^{***} 	&	      	&	 -7.718^{***} 	&	  4.519^{***} 	&	 1.001^{ } 	&	        	&	         	&	         	&	         	&	       	&	       	&	         	&	         	&	         	\\
	&	         	&	    (0.029) 	&	     (0.034) 	&	      	&	      (0.328) 	&	       (0.81) 	&	   (0.838) 	&	        	&	         	&	         	&	         	&	       	&	       	&	         	&	         	&	         	\\
GM--t	&	         	&	 0.067^{**} 	&	 0.926^{***} 	&	      	&	 -7.813^{***} 	&	   10.079^{ } 	&	 1.19^{**} 	&	    8.16^{ } 	&	         	&	         	&	         	&	       	&	       	&	         	&	         	&	         	\\
	&	         	&	    (0.029) 	&	     (0.034) 	&	      	&	       (0.99) 	&	     (21.514) 	&	   (0.583) 	&	    (25.366) 	&	         	&	         	&	         	&	       	&	       	&	         	&	         	&	         	\\
SAV	&	 -0.069^{***} 	&	       	&	        	&	      	&	         	&	         	&	      	&	        	&	 -0.968^{***} 	&	    0.015^{ } 	&	         	&	       	&	       	&	         	&	         	&	 -0.927^{***} 	\\
	&	       (0.01) 	&	       	&	        	&	      	&	         	&	         	&	      	&	        	&	      (0.284) 	&	      (0.066) 	&	         	&	       	&	       	&	         	&	         	&	      (0.149) 	\\
AS	&	        0.000^{ } 	&	       	&	        	&	      	&	         	&	         	&	      	&	        	&	  0.972^{***} 	&	    0.005^{ } 	&	 -0.098^{***} 	&	       	&	       	&	         	&	         	&	 -1.134^{***} 	\\
	&	          (0.000) 	&	       	&	        	&	      	&	         	&	         	&	      	&	        	&	      (0.014) 	&	      (0.027) 	&	      (0.025) 	&	       	&	       	&	         	&	         	&	      (0.144) 	\\
IG	&	        0.000^{ } 	&	       	&	        	&	      	&	         	&	         	&	      	&	        	&	  0.864^{***} 	&	   0.291^{**} 	&	         	&	       	&	       	&	         	&	         	&	 -1.073^{***} 	\\
	&	          (0.000) 	&	       	&	        	&	      	&	         	&	         	&	      	&	        	&	      (0.061) 	&	      (0.113) 	&	         	&	       	&	       	&	         	&	         	&	       (0.39) 	\\
QR	&	 -0.015^{***} 	&	       	&	        	&	      	&	         	&	         	&	      	&	        	&	  -0.215^{**} 	&	 -0.385^{***} 	&	  -0.197^{**} 	&	 -0.134^{ } 	&	 -0.006^{ } 	&	 -0.238^{***} 	&	         	&	 -1.123^{***} 	\\
	&	      (0.002) 	&	       	&	        	&	      	&	         	&	         	&	      	&	        	&	      (0.093) 	&	      (0.094) 	&	      (0.094) 	&	    (0.093) 	&	    (0.083) 	&	      (0.092) 	&	         	&	      (0.141) 	\\
QR--X 	&	 -0.011^{***} 	&	       	&	        	&	      	&	         	&	         	&	      	&	        	&	  -0.225^{**} 	&	 -0.347^{***} 	&	   -0.153^{ } 	&	 -0.108^{ } 	&	 -0.041^{ } 	&	  -0.207^{**} 	&	   -0.548^{*} 	&	 -1.177^{***} 	\\
	&	      (0.004) 	&	       	&	        	&	      	&	         	&	         	&	      	&	        	&	      (0.091) 	&	      (0.095) 	&	      (0.096) 	&	    (0.089) 	&	    (0.093) 	&	      (0.093) 	&	      (0.314) 	&	      (0.143) 	\\
MF--QR	&	  -0.01^{***} 	&	       	&	        	&	      	&	         	&	 -0.461^{***} 	&	       1.2 	&	        	&	 -0.241^{***} 	&	 -0.346^{***} 	&	   -0.079^{ } 	&	 -0.104^{ } 	&	   0.080^{ } 	&	  -0.199^{**} 	&	         	&	 -1.041^{***} 	\\
	&	      (0.002) 	&	       	&	        	&	      	&	         	&	      (0.112) 	&	      	&	        	&	       (0.080) 	&	      (0.089) 	&	      (0.079) 	&	    (0.078) 	&	    (0.073) 	&	      (0.082) 	&	         	&	      (0.136) 	\\
MF--QR--X 	&	    0.002^{ } 	&	       	&	        	&	      	&	         	&	 -0.596^{***} 	&	       1.1 	&	        	&	   -0.228^{*} 	&	 -0.322^{***} 	&	     0.010^{ } 	&	 -0.066^{ } 	&	  0.072^{ } 	&	    -0.19^{*} 	&	 -1.185^{***} 	&	 -1.157^{***} 	\\
	&	      (0.001) 	&	       	&	        	&	      	&	         	&	      (0.115) 	&	      	&	        	&	      (0.117) 	&	      (0.114) 	&	      (0.104) 	&	    (0.114) 	&	    (0.102) 	&	        (0.100) 	&	      (0.086) 	&	      (0.352) 	\\

\bottomrule
\addlinespace
\end{tabular}
\vspace{-0.2cm}
      \begin{tablenotes}[flushleft]
        \setlength\labelsep{0pt}
        \footnotesize
        \item \textbf{Notes}: The table reports the in-sample estimates of the parametric and semi-parametric models (whose functional forms are in Table \ref{tab:models_eq}). To save space, $\beta_0$ and $\zeta$ reported in Table \ref{tab:models_eq} corresponds here to $\omega$ and $\theta$, respectively. Parametric models use Quasi Maximum Likelihood standard errors, semi-parametric models use bootstrap-based standard errors. The weighting parameters of the proposed MF--QR and MF--QR--X models are without the standard errors because they are obtained (and not estimated) via profiling out the weighting parameter $\omega_2$, as described in Section \ref{sec:estimation}. $^{*}$, $^{**}$ and $^{***}$ represent the significance at levels 10\%, 5\% and 1\%, respectively. The sample covers the period from 4 January 2010 to 30 December 2016 (1760 observations). The VaR and ES are calculated at the level $\tau=0.05$. 
      \end{tablenotes}
    \end{threeparttable}
  \end{adjustbox}
  \end{table}
\end{landscape}
\begin{landscape}
\begin{table}
    \centering
\caption{In-sample estimates for Gasoline \label{tab:in_sample_est_gasoline}}
 \vspace{-0.2cm}
\begin{adjustbox}{max height=0.22\textheight}
\begin{threeparttable}
\begin{tabular}{l  ...............}
\toprule
	&	        \mc{$\omega$} 	&	       \alpha 	&	       \beta 	&	     \gamma 	&	            m 	&	      \theta 	&	    \omega_2 	&	          \nu 	&	       \beta_1 	&	      \beta_2 	&	       \beta_3 	&	       \beta_4 	&	       \beta_5 	&	        \beta_X	&	     \gamma_{ES}	\\
\midrule
GARCH--N	&	        0.000^{ } 	&	   0.117^{*} 	&	 0.838^{***} 	&		&	 	&	 	&		&		&	 	&	 	&	 	&	       	&	 	&	 	&	 	\\
	&	(0.000)	&	     (0.062) 	&	     (0.096) 	&		&	 	&	 	&		&		&	 	&	 	&	 	&	       	&	 	&	 	&	 	\\
GARCH--t 	&	        0.000^{ } 	&	 0.038^{***} 	&	 0.955^{***} 	&		&	 	&	 	&		&	 4.702^{***} 	&	 	&	 	&	 	&	       	&	 	&	 	&	 	\\
	&	(0.000)	&	     (0.011) 	&	     (0.014) 	&		&	 	&	 	&		&	     (0.679) 	&	 	&	 	&	 	&	       	&	 	&	 	&	 	\\
GJR--N	&	        0.000^{ } 	&	   0.075^{ } 	&	 0.861^{***} 	&	   0.056^{ } 	&	 	&	 	&		&		&	 	&	 	&	 	&	       	&	 	&	 	&	 	\\
	&	(0.000)	&	     (0.084) 	&	     (0.089) 	&	     (0.063) 	&	 	&	 	&		&		&	 	&	 	&	 	&	       	&	 	&	 	&	 	\\
GJR--t	&	        0.000^{ } 	&	       0.000^{ } 	&	 0.974^{***} 	&	 0.043^{***} 	&	 	&	 	&		&	 4.735^{***} 	&	 	&	 	&	 	&	       	&	 	&	 	&	 	\\
	&	(0.000)	&	     (0.005) 	&	     (0.004) 	&	      (0.01) 	&	 	&	 	&		&	     (0.679) 	&	 	&	 	&	 	&	       	&	 	&	 	&	 	\\
GM--N 	&	 	&	 0.168^{***} 	&	 0.675^{***} 	&		&	 -8.316^{***} 	&	 38.342^{***} 	&	 1.001^{***} 	&		&	 	&	 	&	 	&	       	&	 	&	 	&	 	\\
	&	 	&	     (0.057) 	&	     (0.116) 	&		&	      (0.164) 	&	(3.223)	&	     (0.206) 	&		&	 	&	 	&	 	&	       	&	 	&	 	&	 	\\
GM--t	&	 	&	   0.033^{ } 	&	  0.96^{***} 	&		&	 -7.786^{***} 	&	   10.653^{*} 	&	 1.001^{***} 	&	 4.651^{***} 	&	 	&	 	&	 	&	       	&	 	&	 	&	 	\\
	&	 	&	     (0.093) 	&	     (0.131) 	&		&	      (0.358) 	&	(5.953)	&	     (0.181) 	&	     (0.706) 	&	 	&	 	&	 	&	       	&	 	&	 	&	 	\\
SAV	&	   -0.001^{ } 	&		&		&		&	 	&	 	&		&		&	  0.857^{***} 	&	 -0.227^{***} 	&	 	&	       	&	 	&	 	&	 -0.763^{***} 	\\
	&	      (0.001) 	&		&		&		&	 	&	 	&		&		&	      (0.052) 	&	      (0.068) 	&	 	&	       	&	 	&	 	&	      (0.129) 	\\
AS	&	   -0.001^{ } 	&		&		&		&	 	&	 	&		&		&	  0.859^{***} 	&	  -0.167^{**} 	&	 -0.262^{***} 	&	       	&	 	&	 	&	 -0.718^{***} 	\\
	&	      (0.001) 	&		&		&		&	 	&	 	&		&		&	      (0.055) 	&	       (0.08) 	&	      (0.084) 	&	       	&	 	&	 	&	      (0.136) 	\\
IG	&	        0.000^{ } 	&		&		&		&	 	&	 	&		&		&	  0.754^{***} 	&	   0.509^{**} 	&	 	&	       	&	 	&	 	&	  -0.73^{***} 	\\
	&	(0.000)	&		&		&		&	 	&	 	&		&		&	      (0.098) 	&	       (0.22) 	&	 	&	       	&	 	&	 	&	      (0.251) 	\\
QR	&	 -0.016^{***} 	&		&		&		&	 	&	 	&		&		&	 -0.319^{***} 	&	   -0.207^{*} 	&	    -0.21^{*} 	&	 -0.064^{ } 	&	 -0.316^{***} 	&	 	&	 -0.696^{***} 	\\
	&	      (0.003) 	&		&		&		&	 	&	 	&		&		&	      (0.115) 	&	      (0.116) 	&	      (0.115) 	&	     (0.11) 	&	      (0.121) 	&	 	&	      (0.141) 	\\
QR--X 	&	  -0.013^{**} 	&		&		&		&	 	&	 	&		&		&	 -0.315^{***} 	&	   -0.221^{*} 	&	   -0.225^{*} 	&	 -0.074^{ } 	&	  -0.291^{**} 	&	   -0.315^{ } 	&	 -0.719^{***} 	\\
	&	      (0.005) 	&		&		&		&	 	&	 	&		&		&	      (0.122) 	&	      (0.128) 	&	      (0.115) 	&	    (0.106) 	&	      (0.116) 	&	        (0.4) 	&	      (0.157) 	\\
MF--QR	&	 -0.011^{***} 	&		&		&		&	 	&	 -0.606^{***} 	&	1.200	&		&	 -0.317^{***} 	&	   -0.214^{*} 	&	   -0.137^{ } 	&	 -0.069^{ } 	&	   -0.087^{ } 	&	 	&	 -0.954^{***} 	\\
	&	      (0.002) 	&		&		&		&	 	&	      (0.158) 	&		&		&	      (0.113) 	&	      (0.115) 	&	      (0.108) 	&	    (0.102) 	&	      (0.095) 	&	 	&	      (0.133) 	\\
MF--QR--X	&	   -0.001^{ } 	&		&		&		&	 	&	  -0.762^{**} 	&	1.000	&		&	  -0.272^{**} 	&	   -0.156^{ } 	&	   -0.156^{ } 	&	  -0.04^{ } 	&	   -0.036^{ } 	&	 -0.894^{***} 	&	 -0.941^{***} 	\\
	&	      (0.004) 	&		&		&		&	 	&	      (0.357) 	&		&		&	      (0.106) 	&	      (0.106) 	&	      (0.105) 	&	    (0.098) 	&	      (0.088) 	&	       (0.16) 	&	      (0.133) 	\\

\bottomrule
\addlinespace
\end{tabular}
\vspace{-0.2cm}
      \begin{tablenotes}[flushleft]
        \setlength\labelsep{0pt}
        \footnotesize
        \item \textbf{Notes}: The table reports the in-sample estimates of the parametric and semi-parametric models (whose functional forms are in Table \ref{tab:models_eq}). To save space, $\beta_0$ and $\zeta$ reported in Table \ref{tab:models_eq} corresponds here to $\omega$ and $\theta$, respectively. Parametric models use Quasi Maximum Likelihood standard errors, semi-parametric models use bootstrap-based standard errors. The weighting parameters of the proposed MF--QR and MF--QR--X models are without the standard errors because they are obtained (and not estimated) via profiling out the weighting parameter $\omega_2$, as described in Section \ref{sec:estimation}. $^{*}$, $^{**}$ and $^{***}$ represent the significance at levels 10\%, 5\% and 1\%, respectively. The sample covers the period from 4 January 2010 to 30 December 2016 (1759 observations). The VaR and ES are calculated at the level $\tau=0.05$. 
      \end{tablenotes}
    \end{threeparttable}
  \end{adjustbox}
  \end{table}
\end{landscape}

The in-sample backtesting evaluations are reported in Tables \ref{tab:backtesting_in_sample_crude_oil} (Crude Oil) and \ref{tab:backtesting_in_sample_gasoline} (Gasoline). All models pass the chosen backtesting procedures (p-values in columns 3-7), with a strong preference for the longer windows in the HS non-parametric model.

\begin{table}[htbp]
	\centering
		\caption{In-sample backtesting for Crude Oil}
		\vspace{-0.2cm}
	\label{tab:backtesting_in_sample_crude_oil}
	\begin{adjustbox}{max width=0.75\textwidth}
	\begin{threeparttable}
\begin{tabular}{l  ......}
\toprule
&&\multicolumn{3}{c}{VaR}&\multicolumn{2}{c}{ES}\\
& \mc{AE} & \mc{UC} & \mc{CC} & \mc{DQ} & \mc{UC} & \mc{CC} \\ 
\midrule
GARCH--N & 1.080 & \cellcolor{gray!75}0.449 & \cellcolor{gray!75}0.695 & \cellcolor{gray!75}0.198 & \cellcolor{gray!75}0.058 & \cellcolor{gray!75}0.449 \\ 
  GARCH--t & 1.159 & \cellcolor{gray!75}0.135 & \cellcolor{gray!75}0.327 & \cellcolor{gray!75}0.118 & \cellcolor{gray!75}0.051 & \cellcolor{gray!75}0.135 \\ 
  GJR--N & 1.068 & \cellcolor{gray!75}0.516 & \cellcolor{gray!75}0.733 & \cellcolor{gray!75}0.296 & \cellcolor{gray!75}0.084 & \cellcolor{gray!75}0.516 \\ 
  GJR--t & 1.125 & \cellcolor{gray!75}0.238 & \cellcolor{gray!75}0.489 & \cellcolor{gray!75}0.142 & \cellcolor{gray!75}0.093 & \cellcolor{gray!75}0.238 \\ 
  GM--N & 1.068 & \cellcolor{gray!75}0.516 & \cellcolor{gray!75}0.733 & \cellcolor{gray!75}0.408 & \cellcolor{gray!75}0.06 & \cellcolor{gray!75}0.516 \\ 
  GM--t & 1.136 & \cellcolor{gray!75}0.199 & \cellcolor{gray!75}0.433 & \cellcolor{gray!75}0.254 & \cellcolor{gray!75}0.069 & \cellcolor{gray!75}0.199 \\ 
  HS (w=25) & 1.716 & 0.000 & 0.000 & 0.000 & 0.000 & 0.000 \\ 
  HS (w=50) & 1.398 & 0.000 & 0.001 & 0.000 & 0.000 & 0.000 \\ 
  HS (w=100) & 1.318 & 0.003 & 0.014 & 0.001 & 0.002 & 0.003 \\ 
  HS (w=250) & 1.023 & \cellcolor{gray!75}0.827 & \cellcolor{gray!75}0.688 & 0.000 & \cellcolor{gray!75}0.22 & \cellcolor{gray!75}0.827 \\ 
  HS (w=500) & 1.045 & \cellcolor{gray!75}0.664 & \cellcolor{gray!75}0.84 & 0.000 & \cellcolor{gray!75}0.127 & \cellcolor{gray!75}0.664 \\ 
  SAV & 1.000 & \cellcolor{gray!75}1.000 & \cellcolor{gray!75}0.979 & \cellcolor{gray!75}0.779 & \cellcolor{gray!75}0.458 & \cellcolor{gray!75}1.000 \\ 
  AS & 0.989 & \cellcolor{gray!75}0.913 & \cellcolor{gray!75}0.936 & \cellcolor{gray!75}0.054 & \cellcolor{gray!75}0.488 & \cellcolor{gray!75}0.913 \\ 
  IG & 1.000 & \cellcolor{gray!75}1.000 & \cellcolor{gray!75}0.958 & \cellcolor{gray!75}0.488 & \cellcolor{gray!75}0.439 & \cellcolor{gray!75}1.000 \\ 
  QR & 1.000 & \cellcolor{gray!75}1.000 & \cellcolor{gray!75}0.759 & \cellcolor{gray!75}0.926 & \cellcolor{gray!75}0.463 & \cellcolor{gray!75}1.000 \\ 
  QR--X & 1.000 & \cellcolor{gray!75}1.000 & \cellcolor{gray!75}0.979 & \cellcolor{gray!75}0.983 & \cellcolor{gray!75}0.479 & \cellcolor{gray!75}1.000 \\ 
  MF--QR & 0.989 & \cellcolor{gray!75}0.913 & \cellcolor{gray!75}0.712 & \cellcolor{gray!75}0.925 & \cellcolor{gray!75}0.500 & \cellcolor{gray!75}0.913 \\ 
  MF--QR--X & 1.000 & \cellcolor{gray!75}1.000 & \cellcolor{gray!75}0.958 & \cellcolor{gray!75}0.811 & \cellcolor{gray!75}0.477 & \cellcolor{gray!75}1.000 \\ 
 \bottomrule
\addlinespace
 \end{tabular}
 \vspace{-0.2cm}
	\begin{tablenotes}[flushleft]
   \setlength\labelsep{0pt}
   \linespread{1.25}
\item  \textbf{Notes}:  The table reports the Actual over Expected exceedance  ratio (AE), the p-values of the Unconditional Coverage \citep[UC,][]{Kupiec:1995}, Conditional Coverage \citep[CC,][]{Christoffersen:1998}, Dynamic Quantile \citep[DQ,][]{Engle:Manganelli:2004} tests for the VaR, and the UC and CC tests for ES \citep{Acerbi:Szekely:2014} tests. Dark shade of grey indicates that the model in row passes the test at the 5\% significance level. The sample covers the period from 4 January 2010 to 30 December 2016 (1760 observations). The VaR and ES are calculated at the level $\tau=0.05$.
\end{tablenotes}
\end{threeparttable}
	\end{adjustbox}
\end{table}

\begin{table}[htbp]
	\centering
		\caption{In-sample backtesting for Gasoline}
		\vspace{-0.2cm}
	\label{tab:backtesting_in_sample_gasoline}
	\begin{adjustbox}{max width=0.75\textwidth}
	\begin{threeparttable}
\begin{tabular}{l  ......}
\toprule
&&\multicolumn{3}{c}{VaR}&\multicolumn{2}{c}{ES}\\
& \mc{AE} & \mc{UC} & \mc{CC} & \mc{DQ} & \mc{UC} & \mc{CC} \\ 
\midrule
GARCH--N & 0.955 & \cellcolor{gray!75}0.663 & \cellcolor{gray!75}0.78 & \cellcolor{gray!75}0.994 & \cellcolor{gray!75}0.136 & \cellcolor{gray!75}0.663 \\ 
  GARCH--t & 1.114 & \cellcolor{gray!75}0.28 & \cellcolor{gray!75}0.542 & \cellcolor{gray!75}0.533 & \cellcolor{gray!75}0.088 & \cellcolor{gray!75}0.28 \\ 
  GJR--N & 0.921 & \cellcolor{gray!75}0.441 & \cellcolor{gray!75}0.736 & \cellcolor{gray!75}0.929 & \cellcolor{gray!75}0.198 & \cellcolor{gray!75}0.441 \\ 
  GJR--t & 0.989 & \cellcolor{gray!75}0.917 & \cellcolor{gray!75}0.937 & \cellcolor{gray!75}0.778 & \cellcolor{gray!75}0.28 & \cellcolor{gray!75}0.917 \\ 
  GM--N & 0.966 & \cellcolor{gray!75}0.746 & \cellcolor{gray!75}0.792 & \cellcolor{gray!75}0.995 & \cellcolor{gray!75}0.189 & \cellcolor{gray!75}0.746 \\ 
  GM--t & 1.069 & \cellcolor{gray!75}0.513 & \cellcolor{gray!75}0.73 & \cellcolor{gray!75}0.593 & \cellcolor{gray!75}0.167 & \cellcolor{gray!75}0.513 \\ 
  HS (w=25) & 1.58 & 0.000 & 0.000 & 0.000 & 0.000 & 0.000 \\ 
  HS (w=50) & 1.239 & 0.026 & \cellcolor{gray!75}0.057 & 0.016 & 0.000 & 0.026 \\ 
  HS (w=100) & 1.035 & \cellcolor{gray!75}0.74 & \cellcolor{gray!75}0.889 & \cellcolor{gray!75}0.059 & \cellcolor{gray!75}0.091 & \cellcolor{gray!75}0.74 \\ 
  HS (w=250) & 1.046 & \cellcolor{gray!75}0.66 & \cellcolor{gray!75}0.837 & \cellcolor{gray!75}0.106 & \cellcolor{gray!75}0.149 & \cellcolor{gray!75}0.66 \\ 
  HS (w=500) & 1.001 & \cellcolor{gray!75}0.996 & \cellcolor{gray!75}0.408 & 0.001 & \cellcolor{gray!75}0.272 & \cellcolor{gray!75}0.996 \\ 
  SAV & 1.001 & \cellcolor{gray!75}0.996 & \cellcolor{gray!75}0.758 & \cellcolor{gray!75}0.925 & \cellcolor{gray!75}0.477 & \cellcolor{gray!75}0.996 \\ 
  AS & 1.001 & \cellcolor{gray!75}0.996 & \cellcolor{gray!75}0.958 & \cellcolor{gray!75}0.993 & \cellcolor{gray!75}0.476 & \cellcolor{gray!75}0.996 \\ 
  IG & 1.001 & \cellcolor{gray!75}0.996 & \cellcolor{gray!75}0.979 & \cellcolor{gray!75}0.959 & \cellcolor{gray!75}0.478 & \cellcolor{gray!75}0.996 \\ 
  QR & 1.001 & \cellcolor{gray!75}0.996 & \cellcolor{gray!75}0.958 & \cellcolor{gray!75}0.971 & \cellcolor{gray!75}0.485 & \cellcolor{gray!75}0.996 \\ 
  QR--X & 0.989 & \cellcolor{gray!75}0.917 & \cellcolor{gray!75}0.982 & \cellcolor{gray!75}0.926 & \cellcolor{gray!75}0.5 & \cellcolor{gray!75}0.917 \\ 
  MF--QR & 1.001 & \cellcolor{gray!75}0.996 & \cellcolor{gray!75}0.758 & \cellcolor{gray!75}0.982 & \cellcolor{gray!75}0.478 & \cellcolor{gray!75}0.996 \\ 
  MF--QR--X & 1.001 & \cellcolor{gray!75}0.996 & \cellcolor{gray!75}0.408 & \cellcolor{gray!75}0.817 & \cellcolor{gray!75}0.491 & \cellcolor{gray!75}0.996 \\ 
 \bottomrule
\addlinespace
 \end{tabular}
 \vspace{-0.2cm}
	\begin{tablenotes}[flushleft]
   \setlength\labelsep{0pt}
   \linespread{1.25}
\item  \textbf{Notes}:  The table reports the Actual over Expected exceedance ratio (AE), the p-values of the Unconditional Coverage \citep[UC,][]{Kupiec:1995}, Conditional Coverage \citep[CC,][]{Christoffersen:1998}, Dynamic Quantile \citep[DQ,][]{Engle:Manganelli:2004} tests for the VaR, and the UC and CC tests for ES \citep{Acerbi:Szekely:2014} tests. Dark shade of grey indicates that the model in row passes the test at the 5\% significance level. The sample covers the period from 4 January 2010 to 30 December 2016 (1759 observations). The VaR and ES are calculated at the level $\tau=0.05$.
\end{tablenotes}
\end{threeparttable}
	\end{adjustbox}
\end{table}

\subsection*{Out-of-sample evaluation}

\begin{figure}[htbp]
\caption{MF--QR--X VaR and ES forecasts}
\label{fig:VaR_oos}       
\vspace{-0.2cm}
\includegraphics[width=1\textwidth]{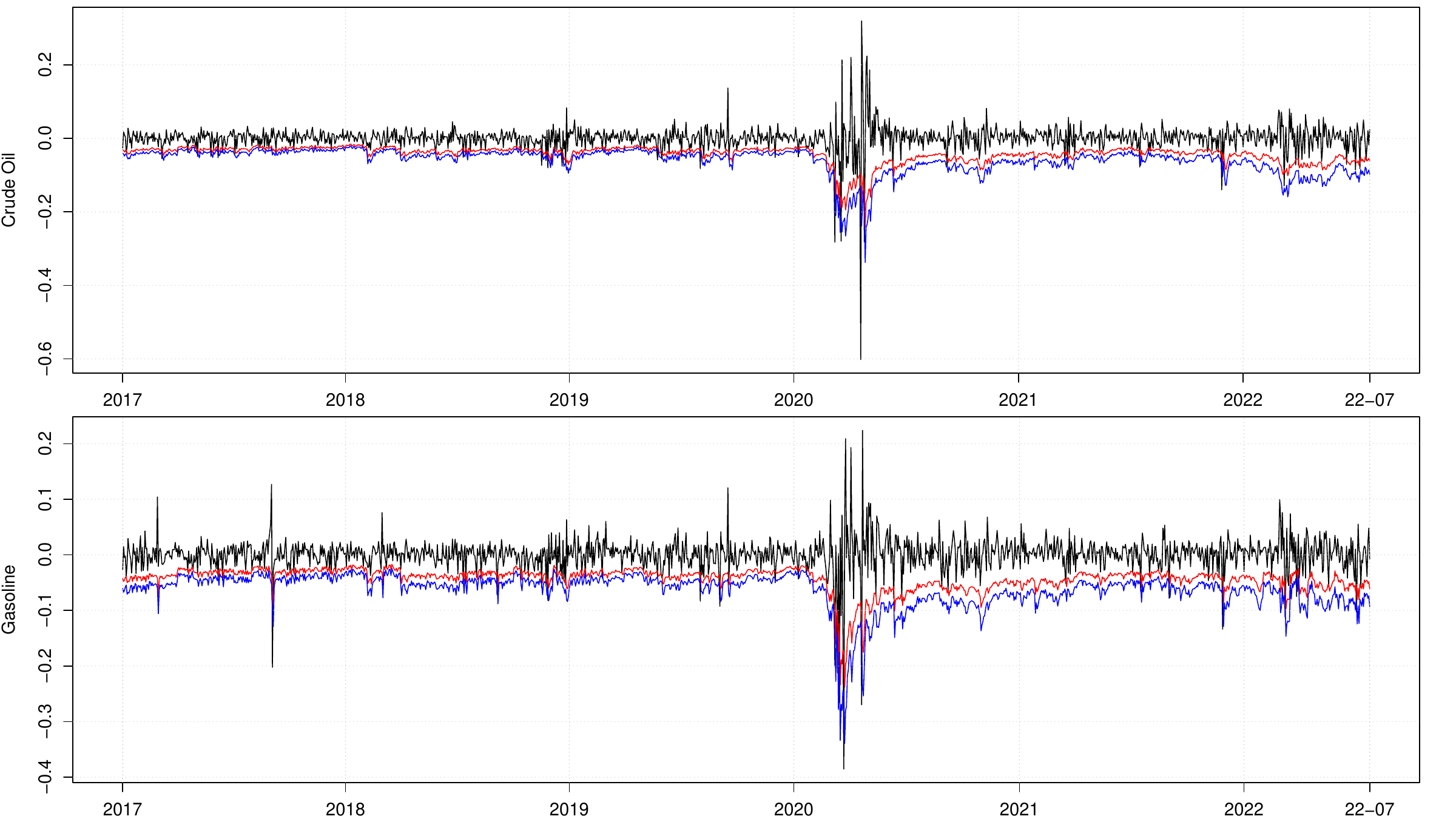}
\begin{minipage}[t]{1\textwidth}
\textbf{Notes:} Plot of the Crude Oil (top) and Gasoline (bottom) daily log-returns (black lines) and of the VaR (red lines) and ES (blue lines) forecasts obtained from the MF--QR--X model.  Sample period: from January 2017  to July 2022.
\end{minipage}
\end{figure}

The empirical analysis is completed by the out-of-sample analysis. In line with \cite{Lazar:Xue:2020}, the one-step-ahead VaR and ES forecasts of the parametric and semi-parametric models are obtained with parameters estimated every five days, using a rolling window of size 1500 observations.  For our main MF--QR--X model, the VaR and ES forecasts are graphically reported in Figure \ref{fig:VaR_oos}.

The results of the out--of--sample evaluations are synthesized in Tables \ref{tab:backtesting_oos_crude_oil} (Crude Oil) and  \ref{tab:backtesting_oos_gasoline} (Gasoline), respectively. \textcolor{black}{While the AE ratios closest  to one are seen for model GM--N for Crude Oil in Table \ref{tab:backtesting_oos_crude_oil}, and for model QR for Gasoline (Table \ref{tab:backtesting_oos_gasoline}), a more formal statistical evaluation of the VaR and ES performances by different models is given by backtesting procedures. C}ontrary to the in-sample period where almost all the models passed the backtesting procedures, going out-of-sample, the proposed MF--QR--X is the only one that \textcolor{black}{fails to reject the null for all the VaR and ES tests} for both the Crude Oil and the Gasoline log-returns (\textcolor{black}{while the QR model passes all  tests only for the latter}), with more scattered and less systematic evidence for the other models, but with a consistent failure of all the tests by GM--t, short window HS and SAV, AS and IG. In Appendix \hyperref[sec:appendix_c]{C}, we also report the results of the backtesting evaluations using a slower frequency (ten/twenty days) of parameter updates. The results are quite robust to different frequency updating schemes, as it can be seen in Tables from \ref{tab:backtesting_oos_crude_oil_10} to \ref{tab:backtesting_oos_gasoline_20}.

\begin{table}[htbp]
	\centering
		\caption{Out-of-sample backtesting for Crude Oil}
		\vspace{-0.2cm}
	\label{tab:backtesting_oos_crude_oil}
\begin{adjustbox}{max width=0.75\textwidth}
	\begin{threeparttable}
\begin{tabular}{l  ......}
\toprule
&&\multicolumn{3}{c}{VaR}&\multicolumn{2}{c}{ES}\\
& \mc{AE} & \mc{UC} & \mc{CC} & \mc{DQ} & \mc{UC} & \mc{CC} \\ 
\midrule
GARCH--N & 1.171 & \cellcolor{gray!75}0.151 & \cellcolor{gray!75}0.356 & 0.018 & 0.000 & \cellcolor{gray!75}0.151 \\ 
  GARCH--t & 1.357 & 0.004 & 0.014 & 0.001 & 0.000 & 0.004 \\ 
  GJR--N & 1.200 & \cellcolor{gray!75}0.096 & \cellcolor{gray!75}0.249 & 0.019 & 0.000 & \cellcolor{gray!75}0.096 \\ 
  GJR--t & 1.314 & 0.01 & 0.032 & 0.001 & 0.000 & 0.01 \\ 
  GM--N & 1.014 & \cellcolor{gray!75}0.903 & \cellcolor{gray!75}0.757 & 0.001 & 0.000 & \cellcolor{gray!75}0.903 \\ 
  GM--t & 1.486 & 0.000 & 0.000 & 0.000 & 0.000 & 0.000 \\ 
  HS (w=25) & 1.671 & 0.000 & 0.000 & 0.000 & 0.000 & 0.000 \\ 
  HS (w=50) & 1.400 & 0.001 & 0.004 & 0.000 & 0.000 & 0.001 \\ 
  HS (w=100) & 1.300 & 0.014 & 0.021 & 0.000 & 0.000 & 0.014 \\ 
  HS (w=250) & 1.229 & \cellcolor{gray!75}0.058 & 0.047 & 0.000 & 0.000 & \cellcolor{gray!75}0.058 \\ 
  HS (w=500) & 1.157 & \cellcolor{gray!75}0.188 & 0.011 & 0.000 & 0.006 & \cellcolor{gray!75}0.188 \\ 
  SAV & 2.543 & 0.000 & 0.000 & 0.000 & 0.000 & 0.000 \\ 
  AS & 3.314 & 0.000 & 0.000 & 0.000 & 0.000 & 0.000 \\ 
  IG & 2.414 & 0.000 & 0.000 & 0.000 & 0.000 & 0.000 \\ 
  QR & 1.086 & \cellcolor{gray!75}0.468 & \cellcolor{gray!75}0.646 & \cellcolor{gray!75}0.844 & 0.006 & \cellcolor{gray!75}0.468 \\ 
  QR--X & 0.900 & \cellcolor{gray!75}0.383 & \cellcolor{gray!75}0.529 & 0.029 & 0.001 & \cellcolor{gray!75}0.317 \\ 
  MF--QR & 1.229 & \cellcolor{gray!75}0.058 & \cellcolor{gray!75}0.136 & \cellcolor{gray!75}0.128 & 0.000 & \cellcolor{gray!75}0.058 \\ 
  MF--QR--X & 0.957 & \cellcolor{gray!75}0.711 & \cellcolor{gray!75}0.701 & \cellcolor{gray!75}0.804 & \cellcolor{gray!75}0.128 & \cellcolor{gray!75}0.711 \\ 
 \bottomrule
\addlinespace
 \end{tabular}
\vspace{-0.2cm}
	\begin{tablenotes}[flushleft]
   \setlength\labelsep{0pt}
   \linespread{1.25}
\item  \textbf{Notes}:    The table reports the Actual over Expected exceedance  ratio (AE), the p-values of the Unconditional Coverage \citep[UC,][]{Kupiec:1995}, Conditional Coverage \citep[CC,][]{Christoffersen:1998}, Dynamic Quantile \citep[DQ,][]{Engle:Manganelli:2004} tests for the VaR, and the UC and CC tests for the ES \citep{Acerbi:Szekely:2014}. Dark shade of grey indicates that the model in row passes the test in column at the 5\% significance level. Models' labels and functional forms are in Table \ref{tab:models_eq}. The sample covers the period from 3 January 2017 to 27 July 2022 (1400 observations). Every model has been refitted once every 5 days. The rolling window used is of 1500 observations. The VaR and ES are calculated at the level $\tau=0.05$.
\end{tablenotes}
\end{threeparttable}
	\end{adjustbox}
\end{table}

\begin{table}[htbp]
	\centering
		\caption{Out-of-sample backtesting for Gasoline}
		\vspace{-0.2cm}
	\label{tab:backtesting_oos_gasoline}
\begin{adjustbox}{max width=0.75\textwidth}
	\begin{threeparttable}
\begin{tabular}{l  ......}
\toprule
&&\multicolumn{3}{c}{VaR}&\multicolumn{2}{c}{ES}\\
& \mc{AE} & \mc{UC} & \mc{CC} & \mc{DQ} & \mc{UC} & \mc{CC} \\ 
\midrule
GARCH--N & 1.057 & \cellcolor{gray!75}0.627 & \cellcolor{gray!75}0.76 & \cellcolor{gray!75}0.958 & 0.018 & \cellcolor{gray!75}0.627 \\ 
  GARCH--t & 1.243 & 0.044 & \cellcolor{gray!75}0.13 & \cellcolor{gray!75}0.133 & 0.009 & 0.044 \\ 
  GJR--N & 1.071 & \cellcolor{gray!75}0.544 & \cellcolor{gray!75}0.735 & \cellcolor{gray!75}0.943 & 0.009 & \cellcolor{gray!75}0.544 \\ 
  GJR--t & 1.157 & \cellcolor{gray!75}0.188 & \cellcolor{gray!75}0.415 & \cellcolor{gray!75}0.657 & 0.016 & \cellcolor{gray!75}0.188 \\ 
  GM--N & 0.800 & \cellcolor{gray!75}0.076 & \cellcolor{gray!75}0.182 & 0.000 & \cellcolor{gray!75}0.348 & \cellcolor{gray!75}0.076 \\ 
  GM--t & 1.571 & 0.000 & 0.000 & 0.000 & 0.000 & 0.000 \\ 
  HS (w=25) & 1.714 & 0.000 & 0.000 & 0.000 & 0.000 & 0.000 \\ 
  HS (w=50) & 1.371 & 0.002 & 0.002 & 0.001 & 0.000 & 0.002 \\ 
  HS (w=100) & 1.200 & \cellcolor{gray!75}0.096 & \cellcolor{gray!75}0.058 & 0.000 & 0.003 & \cellcolor{gray!75}0.096 \\ 
  HS (w=250) & 1.171 & \cellcolor{gray!75}0.151 & \cellcolor{gray!75}0.066 & 0.000 & 0.004 & \cellcolor{gray!75}0.151 \\ 
  HS (w=500) & 1.114 & \cellcolor{gray!75}0.335 & 0.027 & 0.000 & 0.037 & \cellcolor{gray!75}0.335 \\ 
  SAV & 2.186 & 0.000 & 0.000 & 0.000 & 0.000 & 0.000 \\ 
  AS & 2.271 & 0.000 & 0.000 & 0.000 & 0.000 & 0.000 \\ 
  IG & 1.771 & 0.000 & 0.000 & 0.000 & 0.000 & 0.000 \\ 
  QR & 1.000 & \cellcolor{gray!75}1.000 & \cellcolor{gray!75}0.436 & \cellcolor{gray!75}0.801 & \cellcolor{gray!75}0.168 & \cellcolor{gray!75}1.000 \\ 
  QR--X & 0.843 & \cellcolor{gray!75}0.166 & 0.017 & 0.000 & 0.000 & 0.041 \\ 
  MF--QR & 1.071 & \cellcolor{gray!75}0.544 & \cellcolor{gray!75}0.515 & \cellcolor{gray!75}0.207 & 0.017 & \cellcolor{gray!75}0.544 \\ 
  MF--QR--X & 0.971 & \cellcolor{gray!75}0.805 & \cellcolor{gray!75}0.899 & \cellcolor{gray!75}0.596 & \cellcolor{gray!75}0.416 & \cellcolor{gray!75}0.805 \\ 
 \bottomrule
 \addlinespace
 \end{tabular}
\vspace{-0.2cm}
	\begin{tablenotes}[flushleft]
   \setlength\labelsep{0pt}
   \linespread{1.25}
\item  \textbf{Notes}:    The table reports the Actual over Expected exceedance  ratio (AE), the p-values of the Unconditional Coverage \citep[UC,][]{Kupiec:1995}, Conditional Coverage \citep[CC,][]{Christoffersen:1998}, Dynamic Quantile \citep[DQ,][]{Engle:Manganelli:2004} tests for the VaR, and the UC and CC tests for the ES \citep{Acerbi:Szekely:2014}. Dark shade of grey indicates that the model in row passes the test in column at the 5\% significance level. Models' labels and functional forms are in Table \ref{tab:models_eq}. The sample covers the period from 3 January 2017 to 27 July 2022 (1400 observations). Every model has been refitted once every 5 days. The rolling window used is of 1500 observations. The VaR and ES are calculated at the level $\tau=0.05$. 
\end{tablenotes}
\end{threeparttable}
	\end{adjustbox}
\end{table}

\section{Concluding Remarks \label{sec:conc}}

This paper suggested the inclusion of mixed--frequency (MF) components in a quantile regression (QR) approach to VaR and ES estimations, within a dynamic model of volatility with the original introduction of a low-- and a high--frequency (\virg{--X}) components: the outcome was labelled MF--QR--X model. Given its nature of quantile regression, no explicit distribution for the returns is necessary and robustness to outliers in the data is guaranteed.

Starting from the assessment of the  weak stationarity conditions of our semi--parametric MF--QR--X process, we suggested an estimation procedure the performance of which was investigated through an extensive Monte Carlo exercise in finite samples. Overall, we have satisfactory properties of the estimates and the resulting VaR forecasts are robust to some misspecification in the weighting parameter entering the mixed--frequency component. 

Energy commodities -- Crude Oil and Gasoline futures -- take the center stage in the illustration of the empirical performance, both in-- and out--of--sample, of the proposed  MF--QR--X model, contrasting it against several popular parametric, non-parametric and semi-parametric alternatives. The results are encouraging since our model is the only model consistently passing all the VaR and ES backtesting procedures out--of--sample. Our empirical results support the use of MF--QR--X models to exploit the information content of mixed-frequency data in a risk management framework.

Further research may focus on the multivariate extension of the tail risk forecasts, as done by \cite{Torres:Lillo:Laniado:2015}, \cite{DiBernardino:Fernandez:2015}, \cite{Bernardi:Maruotti:Petrella:2017}, and \cite{Raponi:Petrella:2019}, among others. Another interesting point would be the investigation of the performance of the MF--QR--X with an asymmetric term, both for what concerns the daily returns and the low--frequency component, as done by \cite{Amendola:Candila:Gallo:2019}, for instance.

\section*{Declarations}
\textbf{Conflicts of interest} Authors have no conflict of interest.

\bibliographystyle{chicago}
\bibliography{BIBLIO}

\newpage
\appendix
\section*{Appendix A \label{sec:appendix_a}}
\setcounter{equation}{0}
\renewcommand{\theequation}{A.\arabic{equation}}

\subsection*{Proof of Theorem \ref{theorem_1}} \label{sec:theorem_1}

\begin{proof}
Let $\lVert x \rVert_{p}= \left(E|x|^p \right)^{1/p}$, and recall that $MV_{t}$ and $X_{i,t}$ are assumed to be weakly stationary processes. Let ${s}$ be the compact time notation \textit{in lieu of} ${i,t}$, that is, 
$$
s \equiv \sum_{j=1}^{t-1} N_j +i.
$$
Moreover, let $\sigma_{s}=(\beta_0 + \beta_1 |r_{s-1}| + \cdots + \beta_q |r_{s-q}| + \theta |WS_{s-1}| + \beta_X |X_{s-1}|)$. Note that $WS_{t}$, obtained as an affine combination of $\left(MV_{t-1},\cdots,MV_{t-K}\right)$, is weakly stationary.

From the model in (\ref{eq:larch_midas_x}), we can write:
\begin{eqnarray}
\lVert r_{s} \rVert_{p}  &=&  \lVert \sigma_{s} z_{s}\rVert_{p} \nonumber \\ 
&=& \lVert \sigma_{s}\rVert_{p} \cdot \lVert z_{s}\rVert_{p}, \label{eq:prof_theor_1}
\end{eqnarray}
given the independence between $\sigma_{s}$ and $z_{s}$.
For $p=1$, the \emph{right hand side} (RHS) of \eqref{eq:prof_theor_1} is zero, because $z_{s} \overset{i.i.d.}{\sim} (0,1)$. 

Let us now focus on $p=2$; let us replace the second term of the RHS of \eqref{eq:prof_theor_1}, having assumed that $\lVert z_{s}\rVert_{2} = z^* < \infty$:
\begin{eqnarray}
\lVert r_{s} \rVert_{r} &=& z^*\left( E (\beta_0 + \beta_1 |r_{s-1}| + \cdots + \beta_q |r_{s-q}| + \theta |WS_{s-1}| + \beta_X |X_{s-1}|)^2 \right)^{1/2} \\
&\leq &  z^* ( \beta_0 + \beta_1 \lVert r_{s-1} \rVert_{2} + \cdots + \beta_q\lVert r_{s-q} \rVert_{2} + \theta \lVert WS_{s-1} \rVert_{2} + \beta_X \lVert X_{s-1} \rVert_{2}).
\end{eqnarray}
%

Let us now translate this expression in matrix notation. Therefore, let us collect terms in a vector indexed by $s$, that is,
$$\xi_{s}=\left(\lVert r_{s} \rVert_{2},\cdots,
\lVert r_{s-q+1} \rVert_{2}, \lVert WS_{s} \rVert_{2}, \lVert X_{s} \rVert_{2} \right)^{'},$$
and let the $(q+2)\times(q+2)$ dimensional companion matrix $A$, the vectors $b$ and $c$
\[
A=
  \begin{bmatrix}
    z^* \beta_1 & z^* \beta_2 & \cdots & z^* \beta_{q-1}& z^* \beta_q & z^* \theta & z^* \beta_x\\
    1  			& 0 		  & \cdots & 0				&  	0		  & 0		   & 0			\\
    0  			& 1 		  & \cdots & 0				&  	0		  & 0		   & 0			\\
    \vdots		& \vdots	  & \vdots & \vdots			&  	\vdots	  & \vdots	   & \vdots		\\
    0  			& 0 		  & \cdots & 1				&  	0		  & 0		   & 0			\\
    0  			& 0 		  & \cdots & 0				&  	0		  & 0		   & 0			\\
    0  			& 0 		  & \cdots & 0				&  	0		  & 0		   & 0			\\
    0  			& 0 		  & \cdots & 0				&  	0		  & 0		   & 0			\\
  \end{bmatrix}, \quad  b =  \begin{bmatrix}
   z^* \beta_0 \\
   0	\\	
    0	\\	
     \vdots	\\	
      0	\\	
       0	\\	
  0	\\	
  0	\\	
  \end{bmatrix} \quad  \text{and} \quad  c =  \begin{bmatrix}
  0\\
   0	\\	
    0	\\	
     \vdots	\\	
      0	\\	
       0	\\	
  \lVert WS_{s} \rVert_{2}	\\	
   \lVert X_{s} \rVert_{2}	\\
    \end{bmatrix},
\]
where we have made us of the fact that, because of the stationarity of $WS_{s}$ and $X_{s}$, the vector $c$ does not depend on time.
Thus, we have:
\begin{equation}
\xi_{s} \leq A \xi_{s-1} + b + c.
\label{eq:state_space_form}
\end{equation}
%


 Substituting recursively $\xi_{s-1}$ backwards, and letting  $I_{q+2}$ be the identity matrix of size $\left(q+2\right)$,
\begin{align}
\xi_{s} \leq & ~ A \left(A\xi_{s-2}+b + c\right) + b + c\\
\leq & ~ A^2 \xi_{s-2} + Ab  +  b  + A c + c\nonumber\\
 \leq & ~ A^2 \xi_{s-2} + \left(I_{q+2} + A\right)b + \left(I_{q+2} + A\right)c\\
 \leq & ~ A^3 \xi_{s-3} + \left(I_{q+2} + A + A^2\right)b + \left(I_{q+2} + A+A^2\right)c\\
 \vdots \nonumber \\
 \leq&~ A^m \xi_{s-m} + \left(I_{q+2}+A+A^2+\cdots+A^{m-1}\right)b + \left(I_{q+2}+A+A^2+\cdots+A^{m-1}\right)c.
\label{eq:lag_m_theorem}
\end{align} 

Recall the characteristic polynomial of $A$ is $\phi(\lambda)$, defined by Eq. (\ref{eq:pol}), namely,
	\begin{equation}
	\phi(\lambda)=z^* \left( \beta_1 \lambda^{q+1} + \beta_2 \lambda^{q} + \cdots + \beta_q \lambda^{q-2}\right) - \lambda^{q+2},
\end{equation}
which has all eigenvalues $\lambda$ lie inside the unit circle. When $m \rightarrow \infty$, for the eigen--decomposition theorem, this implies that
\begin{equation}
\lim_{m\rightarrow \infty} A^m =0,
\end{equation}
and that 
\begin{equation}
\lim_{m\rightarrow \infty} (I_{q+2}+A+A^2+\cdots+A^{m-1}) =(I_{q+2}-A)^{-1}.
\end{equation}
Putting terms together, therefore, as $m\rightarrow \infty$ we can say that 
\begin{equation}
\xi_{s} \leq \left(I_{q+2}-A\right)^{-1}b + \left(I_{q+2}-A\right)^{-1}c<\infty,
\label{eq:final_term}
\end{equation}
that is the RHS converges to a finite expression not depending on time, establishing the result.
\end{proof}

\newpage
\section*{Appendix B\label{sec:appendix_b}}

\textcolor{black}{In what follows, we illustrate the bootstrap procedure used to calculate the standard errors. For simplicity, we focus on the QR model with just only one lag, being the procedure easily extensible to the other semi-parametric models. Let $\widehat{\Theta}_\tau=\left(\hat{\beta}_{0,\tau},\hat{\beta}_{1,\tau}\right)$ be the estimated vector of parameters for the QR model. The resulting VaR is then $\widehat{Q}_{r_{i,t}}\left(\tau\right)$. Letting $r_{i,t}^{(boot)}$ be the bootstrap returns, we assume that $r_{1,1}^{(boot)}=r_{1,1}$. The step-by-step procedure to obtain the bootstrap standard errors is as follows:
\begin{enumerate}
\item Obtain the standardized residuals as $\hat{z}_{i,t}=r_{i,t}/|\widehat{Q}_{r_{i,t}}\left(\tau\right)|$, for all $i$ and $t$.
\item Sample with replacement from $\hat{z}_{i,t}$, obtaining the bootstrap residuals $\hat{z}_{i,t}^{(boot)}$.
\item Obtain the bootstrap series of VaR as $\widehat{Q}_{r_{i,t}^{(boot)}}=\hat{\beta}_{0,\tau}+\hat{\beta}_{1,\tau}r_{i-1,t}^{(boot)}$.
\item Obtain the bootstrap series of returns as $r_{i,t}^{(boot)}=|\widehat{Q}_{r_{i,t}^{(boot)}}|\hat{z}_{i,t}^{(boot)}$.
\item Repeat 2-4 for all $i$ and $t$ to get one complete bootstrap series of $r_{i,t}^{(boot)}$.
\item Estimate the VaR using $r_{i,t}^{(boot)}$, obtaining $\hat{\beta}_{0,\tau}^{(boot)}$ and $\hat{\beta}_{1,\tau}^{(boot)}$.
\item Repeat steps 2-6 $BOOT$ number of times, obtaining the bootstrap series $\left\lbrace\hat{\beta}_{0,\tau}\right\rbrace_{boot=1}^{BOOT}$ and $\left\lbrace\hat{\beta}_{1,\tau}\right\rbrace_{boot=1}^{BOOT}$.  
\end{enumerate}
The bootstrap standard errors for $\hat{\beta}_{0,\tau}$ and $\hat{\beta}_{1,\tau}$ are then obtained as sample standard deviations of the series $\left\lbrace\hat{\beta}_{0,\tau}\right\rbrace_{boot=1}^{BOOT}$ and $\left\lbrace\hat{\beta}_{1,\tau}\right\rbrace_{boot=1}^{BOOT}$, respectively. It is worth noting that the previous procedure can be naturally extended to the models dedicated to the joint estimation of VaR and ES measures.
}
\newpage
\section*{Appendix C\label{sec:appendix_c}}

\begin{table}[htbp]
	\centering
		\caption{Out-of-sample backtesting for Crude Oil. Re-fitting period: 10 days}
		\vspace{-0.2cm}
	\label{tab:backtesting_oos_crude_oil_10}
\begin{adjustbox}{max width=0.75\textwidth}
	\begin{threeparttable}
\begin{tabular}{l  ......}
\toprule
&&\multicolumn{3}{c}{VaR}&\multicolumn{2}{c}{ES}\\
& \mc{AE} & \mc{UC} & \mc{CC} & \mc{DQ} & \mc{UC} & \mc{CC} \\ 
\midrule
GARCH--N & 1.157 & \cellcolor{gray!75}0.188 & \cellcolor{gray!75}0.415 & 0.017 & 0.000 & \cellcolor{gray!75}0.188 \\ 
  GARCH--t & 1.357 & 0.004 & 0.014 & 0.001 & 0.000 & 0.004 \\ 
  GJR--N & 1.200 & \cellcolor{gray!75}0.096 & \cellcolor{gray!75}0.249 & 0.019 & 0.000 & \cellcolor{gray!75}0.096 \\ 
  GJR--t & 1.314 & 0.01 & 0.032 & 0.001 & 0.000 & 0.010 \\ 
  GM--N & 1.014 & \cellcolor{gray!75}0.903 & \cellcolor{gray!75}0.97 & 0.020 & 0.004 & \cellcolor{gray!75}0.903 \\ 
  GM--t & 1.329 & 0.007 & 0.027 & 0.000 & 0.000 & 0.007 \\ 
  HS (w=25) & 1.671 & 0.000 & 0.000 & 0.000 & 0.000 & 0.000 \\ 
  HS (w=50) & 1.400 & 0.001 & 0.004 & 0.000 & 0.000 & 0.001 \\ 
  HS (w=100) & 1.300 & 0.014 & 0.021 & 0.000 & 0.000 & 0.014 \\ 
  HS (w=250) & 1.229 & \cellcolor{gray!75}0.058 & 0.047 & 0.000 & 0.000 & \cellcolor{gray!75}0.058 \\ 
  HS (w=500) & 1.157 & \cellcolor{gray!75}0.188 & 0.011 & 0.000 & 0.006 & \cellcolor{gray!75}0.188 \\ 
  SAV & 1.286 & 0.019 & 0.01 & 0.004 & 0.01 & 0.019 \\ 
  AS & 1.886 & 0.000 & 0.000 & 0.000 & 0.000 & 0.000 \\ 
  IG & 1.429 & 0.001 & 0.000 & 0.000 & 0.002 & 0.001 \\ 
  QR & 1.071 & \cellcolor{gray!75}0.544 & \cellcolor{gray!75}0.722 & \cellcolor{gray!75}0.915 & 0.012 & \cellcolor{gray!75}0.544 \\ 
  QR--X & 0.971 & \cellcolor{gray!75}0.805 & \cellcolor{gray!75}0.154 & 0.000 & 0.000 & \cellcolor{gray!75}0.621 \\ 
  MF--QR & 1.200 & \cellcolor{gray!75}0.096 & \cellcolor{gray!75}0.219 & \cellcolor{gray!75}0.581 & 0.000 & \cellcolor{gray!75}0.096 \\ 
  MF--QR--X & 0.986 & \cellcolor{gray!75}0.902 & \cellcolor{gray!75}0.685 & \cellcolor{gray!75}0.563 & \cellcolor{gray!75}0.085 & \cellcolor{gray!75}0.902 \\ 
 
 \bottomrule
\addlinespace
 \end{tabular}
 \vspace{-0.2cm}
	\begin{tablenotes}[flushleft]
   \setlength\labelsep{0pt}
   \linespread{1.25}
\item  \textbf{Notes}:      The table reports the Actual over Expected exceedance  ratio (AE), the p-values of the Unconditional Coverage \citep[UC,][]{Kupiec:1995}, Conditional Coverage \citep[CC,][]{Christoffersen:1998}, Dynamic Quantile \citep[DQ,][]{Engle:Manganelli:2004} tests for the VaR, and the UC and CC tests for ES \citep{Acerbi:Szekely:2014}. Dark shade of grey indicates that the model in row passes the test at the 5\% significance level. Models' labels and functional forms are in Table \ref{tab:models_eq}. The sample covers the period from 3 January 2017 to 27 July 2022 (1400 observations). Every model has been refitted once every 10 days. The rolling window used is of 1500 observations. The VaR and ES are calculated at the level $\tau=0.05$.
\end{tablenotes}
\end{threeparttable}
	\end{adjustbox}
\end{table}

\begin{table}[htbp]
	\centering
		\caption{Out-of-sample backtesting for Crude Oil. Re-fitting period: 20 days}
		\vspace{-0.2cm}
	\label{tab:backtesting_oos_crude_oil_20}
\begin{adjustbox}{max width=0.75\textwidth}
	\begin{threeparttable}
\begin{tabular}{l  ......}
\toprule
&&\multicolumn{3}{c}{VaR}&\multicolumn{2}{c}{ES}\\
& \mc{AE} & \mc{UC} & \mc{CC} & \mc{DQ} & \mc{UC} & \mc{CC} \\ 
\midrule
GARCH--N & 1.143 & \cellcolor{gray!75}0.23 & \cellcolor{gray!75}0.476 & 0.015 & 0.000 & \cellcolor{gray!75}0.230 \\ 
  GARCH--t & 1.329 & 0.007 & 0.027 & 0.001 & 0.000 & 0.007 \\ 
  GJR--N & 1.186 & \cellcolor{gray!75}0.121 & \cellcolor{gray!75}0.300 & 0.016 & 0.000 & \cellcolor{gray!75}0.121 \\ 
  GJR--t & 1.314 & 0.010 & 0.032 & 0.001 & 0.000 & 0.010 \\ 
  GM--N & 1.057 & \cellcolor{gray!75}0.627 & \cellcolor{gray!75}0.517 & \cellcolor{gray!75}0.058 & 0.001 & \cellcolor{gray!75}0.627 \\ 
  GM--t & 1.257 & 0.033 & \cellcolor{gray!75}0.101 & \cellcolor{gray!75}0.073 & 0.000 & 0.033 \\ 
  HS (w=25) & 1.671 & 0.000 & 0.000 & 0.000 & 0.000 & 0.000 \\ 
  HS (w=50) & 1.400 & 0.001 & 0.004 & 0.000 & 0.000 & 0.001 \\ 
  HS (w=100) & 1.300 & 0.014 & 0.021 & 0.000 & 0.000 & 0.014 \\ 
  HS (w=250) & 1.229 & \cellcolor{gray!75}0.058 & 0.047 & 0.000 & 0.000 & \cellcolor{gray!75}0.058 \\ 
  HS (w=500) & 1.157 & \cellcolor{gray!75}0.188 & 0.011 & 0.000 & 0.006 & \cellcolor{gray!75}0.188 \\ 
  SAV & 1.114 & \cellcolor{gray!75}0.335 & \cellcolor{gray!75}0.263 & \cellcolor{gray!75}0.149 & \cellcolor{gray!75}0.126 & \cellcolor{gray!75}0.335 \\ 
  AS & 1.543 & 0.000 & 0.000 & 0.000 & 0.000 & 0.000 \\ 
  IG & 1.114 & \cellcolor{gray!75}0.335 & \cellcolor{gray!75}0.263 & \cellcolor{gray!75}0.582 & \cellcolor{gray!75}0.084 & \cellcolor{gray!75}0.335 \\ 
 
  QR & 1.086 & \cellcolor{gray!75}0.468 & \cellcolor{gray!75}0.646 & \cellcolor{gray!75}0.853 & 0.007 & \cellcolor{gray!75}0.468 \\ 
  QR--X & 1.043 & \cellcolor{gray!75}0.715 & 0.001 & 0.000 & 0.000 & \cellcolor{gray!75}0.535 \\ 
  MF--QR & 1.200 & \cellcolor{gray!75}0.096 & \cellcolor{gray!75}0.219 & \cellcolor{gray!75}0.509 & 0.000 & \cellcolor{gray!75}0.096 \\ 
  MF--QR--X & 0.971 & \cellcolor{gray!75}0.805 & \cellcolor{gray!75}0.699 & \cellcolor{gray!75}0.521 & \cellcolor{gray!75}0.091 & \cellcolor{gray!75}0.805 \\ 
 
 \bottomrule
 \addlinespace
 \end{tabular}
 \vspace{-0.2cm}
	\begin{tablenotes}[flushleft]
   \setlength\labelsep{0pt}
   \linespread{1.25}
\item  \textbf{Notes}:     The table reports the Actual over Expected exceedance  ratio (AE), the p-values of the Unconditional Coverage \citep[UC,][]{Kupiec:1995}, Conditional Coverage \citep[CC,][]{Christoffersen:1998}, Dynamic Quantile \citep[DQ,][]{Engle:Manganelli:2004} tests for the VaR, and the UC and CC tests for ES \citep{Acerbi:Szekely:2014}. Dark shade of grey indicates that the model in row passes the test at the 5\% significance level. Models' labels and functional forms are in Table \ref{tab:models_eq}. The sample covers the period from 3 January 2017 to 27 July 2022 (1400 observations). Every model has been refitted once every 20 days. The rolling window used is of 1500 observations. The VaR and ES are calculated at the level $\tau=0.05$.
\end{tablenotes}
\end{threeparttable}
	\end{adjustbox}
\end{table}

\begin{table}[htbp]
	\centering
		\caption{Out-of-sample backtesting for Gasoline. Re-fitting period: 10 days}
		\vspace{-0.2cm}
	\label{tab:backtesting_oos_gasoline_10}
\begin{adjustbox}{max width=0.75\textwidth}
	\begin{threeparttable}
\begin{tabular}{l  ......}
\toprule
&&\multicolumn{3}{c}{VaR}&\multicolumn{2}{c}{ES}\\
& \mc{AE} & \mc{UC} & \mc{CC} & \mc{DQ} & \mc{UC} & \mc{CC} \\ 
\midrule
GARCH--N & 1.057 & \cellcolor{gray!75}0.627 & \cellcolor{gray!75}0.76 & \cellcolor{gray!75}0.958 & 0.018 & \cellcolor{gray!75}0.627 \\ 
  GARCH--t & 1.257 & 0.033 & \cellcolor{gray!75}0.102 & \cellcolor{gray!75}0.088 & 0.007 & 0.033 \\ 
  GJR--N & 1.071 & \cellcolor{gray!75}0.544 & \cellcolor{gray!75}0.735 & \cellcolor{gray!75}0.943 & 0.009 & \cellcolor{gray!75}0.544 \\ 
  GJR--t & 1.157 & \cellcolor{gray!75}0.188 & \cellcolor{gray!75}0.415 & \cellcolor{gray!75}0.655 & 0.016 & \cellcolor{gray!75}0.188 \\ 
  GM--N & 0.914 & \cellcolor{gray!75}0.456 & \cellcolor{gray!75}0.385 & 0.01 & \cellcolor{gray!75}0.098 & \cellcolor{gray!75}0.456 \\ 
  GM--t & 1.357 & 0.004 & 0.001 & 0.000 & 0.000 & 0.004 \\ 
  HS (w=25) & 1.714 & 0.000 & 0.000 & 0.000 & 0.000 & 0.000 \\ 
  HS (w=50) & 1.371 & 0.002 & 0.002 & 0.001 & 0.000 & 0.002 \\ 
  HS (w=100) & 1.2 & \cellcolor{gray!75}0.096 & \cellcolor{gray!75}0.058 & 0.000 & 0.003 & \cellcolor{gray!75}0.096 \\ 
  HS (w=250) & 1.171 & \cellcolor{gray!75}0.151 & \cellcolor{gray!75}0.066 & 0.000 & 0.004 & \cellcolor{gray!75}0.151 \\ 
  HS (w=500) & 1.114 & \cellcolor{gray!75}0.335 & 0.027 & 0.000 & 0.037 & \cellcolor{gray!75}0.335 \\ 
  SAV & 1.343 & 0.005 & 0.002 & 0.008 & 0.010 & 0.005 \\ 
  AS & 1.343 & 0.005 & 0.002 & 0.001 & 0.016 & 0.005 \\ 
  IG & 1.214 & \cellcolor{gray!75}0.075 & \cellcolor{gray!75}0.174 & \cellcolor{gray!75}0.129 & \cellcolor{gray!75}0.088 & \cellcolor{gray!75}0.075 \\ 
  QR & 1.029 & \cellcolor{gray!75}0.807 & \cellcolor{gray!75}0.493 & \cellcolor{gray!75}0.706 & \cellcolor{gray!75}0.108 & \cellcolor{gray!75}0.807 \\ 
  QR--X & 0.886 & \cellcolor{gray!75}0.317 & 0.000 & 0.000 & 0.000 & \cellcolor{gray!75}0.100 \\ 
  MF--QR & 1.043 & \cellcolor{gray!75}0.715 & \cellcolor{gray!75}0.510 & \cellcolor{gray!75}0.237 & 0.022 & \cellcolor{gray!75}0.715 \\ 
  MF--QR--X & 0.986 & \cellcolor{gray!75}0.902 & \cellcolor{gray!75}0.939 & \cellcolor{gray!75}0.421 & \cellcolor{gray!75}0.380 & \cellcolor{gray!75}0.902 \\ 
  \bottomrule
  \addlinespace
  \end{tabular}
\vspace{-0.2cm}
	\begin{tablenotes}[flushleft]
   \setlength\labelsep{0pt}
   \linespread{1.25}
\item  \textbf{Notes}:    The table reports the Actual over Expected exceedance  ratio (AE), the p-values of the Unconditional Coverage \citep[UC,][]{Kupiec:1995}, Conditional Coverage \citep[CC,][]{Christoffersen:1998}, Dynamic Quantile \citep[DQ,][]{Engle:Manganelli:2004} tests for the VaR, and the UC and CC tests for ES \citep{Acerbi:Szekely:2014}. Dark shade of grey indicates that the model in row passes the test at the 5\% significance level. Models' labels and functional forms are in Table \ref{tab:models_eq}. The sample covers the period from 3 January 2017 to 27 July 2022 (1400 observations). Every model has been refitted once every 10 days. The rolling window used is of 1500 observations. The VaR and ES are calculated at the level $\tau=0.05$.
\end{tablenotes}
\end{threeparttable}
	\end{adjustbox}
\end{table}

\begin{table}[htbp]
	\centering
		\caption{Out-of-sample backtesting for Gasoline. Re-fitting period: 20 days}
		\vspace{-0.2cm}
	\label{tab:backtesting_oos_gasoline_20}
\begin{adjustbox}{max width=0.75\textwidth}
	\begin{threeparttable}
\begin{tabular}{l  ......}
\toprule
&&\multicolumn{3}{c}{VaR}&\multicolumn{2}{c}{ES}\\
& \mc{AE} & \mc{UC} & \mc{CC} & \mc{DQ} & \mc{UC} & \mc{CC} \\ 
\midrule
GARCH--N & 1.057 & \cellcolor{gray!75}0.627 & \cellcolor{gray!75}0.76 & \cellcolor{gray!75}0.958 & 0.018 & \cellcolor{gray!75}0.627 \\ 
  GARCH--t & 1.243 & 0.044 & \cellcolor{gray!75}0.128 & \cellcolor{gray!75}0.101 & 0.009 & 0.044 \\ 
  GJR--N & 1.071 & \cellcolor{gray!75}0.544 & \cellcolor{gray!75}0.735 & \cellcolor{gray!75}0.943 & 0.009 & \cellcolor{gray!75}0.544 \\ 
  GJR--t & 1.157 & \cellcolor{gray!75}0.188 & \cellcolor{gray!75}0.415 & \cellcolor{gray!75}0.652 & 0.015 & \cellcolor{gray!75}0.188 \\ 
  GM--N & 1.043 & \cellcolor{gray!75}0.715 & \cellcolor{gray!75}0.51 & \cellcolor{gray!75}0.357 & 0.01 & \cellcolor{gray!75}0.715 \\ 
  GM--t & 1.429 & 0.001 & 0.000 & 0.000 & 0.000 & 0.001 \\ 
  HS (w=25) & 1.714 & 0.000 & 0.000 & 0.000 & 0.000 & 0.000 \\ 
  HS (w=50) & 1.371 & 0.002 & 0.002 & 0.001 & 0.000 & 0.002 \\ 
  HS (w=100) & 1.200 & \cellcolor{gray!75}0.096 & \cellcolor{gray!75}0.058 & 0.000 & 0.003 & \cellcolor{gray!75}0.096 \\ 
  HS (w=250) & 1.171 & \cellcolor{gray!75}0.151 & \cellcolor{gray!75}0.066 & 0.000 & 0.004 & \cellcolor{gray!75}0.151 \\ 
  HS (w=500) & 1.114 & \cellcolor{gray!75}0.335 & 0.027 & 0.000 & 0.037 & \cellcolor{gray!75}0.335 \\ 
  SAV & 1.329 & 0.007 & 0.009 & 0.000 & 0.011 & \cellcolor{gray!75}0.23 \\ 
  AS & 1.171 & \cellcolor{gray!75}0.151 & \cellcolor{gray!75}0.231 & \cellcolor{gray!75}0.349 & \cellcolor{gray!75}0.068 & \cellcolor{gray!75}0.151 \\ 
  IG & 1.086 & \cellcolor{gray!75}0.468 & \cellcolor{gray!75}0.697 & \cellcolor{gray!75}0.854 & \cellcolor{gray!75}0.162 & \cellcolor{gray!75}0.468 \\ 
  QR & 1.014 & \cellcolor{gray!75}0.903 & \cellcolor{gray!75}0.468 & \cellcolor{gray!75}0.659 & \cellcolor{gray!75}0.128 & \cellcolor{gray!75}0.903 \\ 
  QR--X & 1.057 & \cellcolor{gray!75}0.627 & 0.000 & 0.000 & 0.000 & \cellcolor{gray!75}0.456 \\ 
  MF--QR & 1.057 & \cellcolor{gray!75}0.627 & \cellcolor{gray!75}0.517 & 0.023 & 0.017 & \cellcolor{gray!75}0.627 \\ 
  MF--QR--X & 0.957 & \cellcolor{gray!75}0.711 & \cellcolor{gray!75}0.844 & \cellcolor{gray!75}0.544 & \cellcolor{gray!75}0.428 & \cellcolor{gray!75}0.711 \\ 
 \bottomrule
\addlinespace
\end{tabular}
\vspace{-0.2cm}
	\begin{tablenotes}[flushleft]
   \setlength\labelsep{0pt}
   \linespread{1.25}
\item  \textbf{Notes}: The table reports the Actual over Expected exceedance  ratio (AE), the p-values of the Unconditional Coverage \citep[UC,][]{Kupiec:1995}, Conditional Coverage \citep[CC,][]{Christoffersen:1998}, Dynamic Quantile \citep[DQ,][]{Engle:Manganelli:2004} tests for the VaR, and the UC and CC tests for ES \citep{Acerbi:Szekely:2014}. Dark shade of grey indicates that the model in row passes the test at the 5\% significance level. Models' labels and functional forms are in Table \ref{tab:models_eq}. The sample covers the period from 3 January 2017 to 27 July 2022 (1400 observations). Every model has been refitted once every 20 days. The rolling window used is of 1500 observations. The VaR and ES are calculated at the level $\tau=0.05$.
\end{tablenotes}
\end{threeparttable}
	\end{adjustbox}
\end{table}

\end{document}